\documentclass[iop]{emulateapj}

\usepackage{amsmath}
\usepackage{hyperref}

\newcommand{\myemail}{mkromer@mpa-garching.mpg.de}

\shorttitle{Synthetic observables for sub-Chandrasekhar-mass models of SNe Ia}
\shortauthors{Kromer et al.}

\begin{document}


\title{Double-detonation sub-Chandrasekhar supernovae: synthetic observables for minimum helium shell mass models}


\author{M.~Kromer, S.~A.~Sim, M.~Fink, F.~K.~R\"opke, I.~R.~Seitenzahl and W.~Hillebrandt}
\affil{Max-Planck-Institut f\"ur Astrophysik, Karl-Schwarzschild-Stra{\ss}e 1, D-85748 Garching b. M\"unchen, Germany}
\email{\myemail}


\begin{abstract}
  In the double detonation scenario for Type Ia supernovae it is suggested 
  that a detonation initiates in a shell of helium-rich material accreted 
  from a companion star by a sub-Chandrasekhar-mass White Dwarf. This shell 
  detonation drives a shock front into the carbon-oxygen White Dwarf that 
  triggers a secondary detonation in the core. The core detonation results 
  in a complete disruption of the White Dwarf. Earlier studies concluded that
  this scenario has difficulties in accounting for the observed properties 
  of Type Ia supernovae since the explosion ejecta are surrounded by the 
  products of explosive helium burning in the shell. Recently, however, it 
  was proposed that detonations might be possible for much less massive 
  helium shells than previously assumed \citep{Bildsten2007a}. Moreover,
  it was shown that even detonations of these minimum helium shell masses 
  robustly trigger detonations of the carbon-oxygen core \citep{Fink2010a}.
  Therefore it is possible that the impact of the helium layer on observables
  is less than previously thought. Here, we present time-dependent 
  multi-wavelength radiative transfer calculations for models with minimum 
  helium shell mass and derive synthetic observables for both the optical 
  and $\gamma$-ray spectral regions. These differ strongly from those found 
  in earlier simulations of sub-Chandrasekhar-mass explosions in which more 
  massive helium shells were considered. Our models predict light curves 
  which cover both the range of brightnesses and the rise and decline times
  of observed Type Ia supernovae. However, their colours and spectra do not 
  match the observations. In particular, their $B-V$ colours are generally 
  too red. We show that this discrepancy is mainly due to the composition of 
  the burning products of the helium shell of the \citet{Fink2010a} models
  which contain significant amounts of titanium and chromium. Using a toy 
  model, we also show that the burning products of the helium shell depend 
  crucially on its initial composition. This leads us to conclude that 
  good agreement between sub-Chandrasekhar-mass explosions and observed 
  Type Ia supernovae may still be feasible but further study of the shell
  properties is required.
\end{abstract}





\section{Introduction} 

Although the standard single-degenerate (SD) Chandrasekhar-mass 
scenario (see \citealt{Hillebrandt2000a} for a review) is capable 
of explaining most of the observed diversity of Type Ia supernovae 
(SNe~Ia) \citep{Hoeflich1996b,Kasen2009a} via the delayed-detonation 
\citep{Khokhlov1991a} model, it suffers from severe problems in 
explaining the observed rate of SNe~Ia. In particular, binary evolution 
population synthesis calculations predict rates which are an order 
of magnitude too low compared to the observed rate of SNe~Ia 
(\citealt{Ruiter2009a}, but see \citealt{Meng2010a}). In addition, 
recent observational studies suggest that Chandrasekhar-mass 
explosions of hydrogen-accreting carbon-oxygen (C/O) White Dwarfs 
(WDs) in SD binary systems cannot account for all SNe~Ia: 
\citet{Gilfanov2010a} found that the X-ray flux of nearby elliptical 
galaxies is significantly weaker than expected for a population of WDs 
accreting hydrogen towards the Chandrasekhar-mass needed to explain 
the observed supernova rate in elliptical galaxies (see also 
\citealt{diStefano2010a}). Moreover, there is growing observational 
evidence that there are different populations of SNe~Ia 
\citep{Mannucci2005a,Scannapieco2005a}.

This has led to a revived interest in alternative explosion mechanisms.
Here we consider the double detonation scenario applied to 
sub-Chandrasekhar-mass C/O WDs. In that scenario a helium-accreting 
C/O WD explodes below $M_\mathrm{Ch}$ due to a detonation in the 
accreted helium shell which triggers a secondary core detonation by 
compressional heating \citep{Woosley1994a,Livne1995a,Fink2007a}. This 
model has some very appealing features. Depending on the initial mass 
of the WD, a wide range of explosion strengths can be realized 
(e.g. \citealt{Woosley1994a,Livne1995a,Hoeflich1996b}). Moreover, 
population synthesis studies \citep{Ruiter2009a} predict rates 
comparable to the observed galactic supernova rate. 

However, earlier work \citep{Hoeflich1996b,Hoeflich1996c,Nugent1997a} 
found light curves and spectra of such models to be in conflict with the 
observed spectra and light curves of SNe~Ia. The differences were
mainly attributed to the composition of the outer layers. Due to the
initial helium detonation in the outer shell, the ejecta of 
sub-Chandrasekhar-mass models are surrounded by a layer of helium 
and its burning products (which can include iron-peak nuclei). This, 
however, is in apparent contradiction to the layered composition 
structure of observed SNe~Ia, where the composition changes from 
iron-group elements in the core to lower mass elements in the outer 
layers. 

In a preceding paper \citet{Sim2010a} have shown that 
artificial explosions of ``naked'' sub-Chandrasekhar-mass WDs can
reproduce the observed diversity of SNe~Ia. Thus it is natural to
ask if somewhat modified properties in the initial helium shells of 
realistic sub-Chandrasekhar-mass models can reduce the negative 
post-explosion effect of this shell on the observables. In particular, 
\citet{Bildsten2007a} recently presented new calculations, indicating 
that detonations might occur for much less massive helium shells than
previously thought. \citet{Fink2010a} adopted the minimum helium shell 
masses of \citet{Bildsten2007a} and investigated whether such low-mass 
helium detonations are capable of triggering a secondary detonation in 
the C/O core of the WD\@. In that study, they concluded that as soon as 
a detonation in the helium shell initiates, a subsequent core detonation 
is virtually inevitable. For example, they found that even a helium 
shell mass as low as $0.039\,M_\odot$ is sufficient to detonate a 
C/O WD of $1.125\,M_\odot$. 

Here, we focus on the observable properties of the models presented in 
\citet{Fink2010a} and their comparison to real SNe Ia. In particular, 
we investigate whether the low helium shell masses of these models help 
to alleviate the problems encountered previously when comparing double 
detonation sub-Chandrasekhar-mass models to observed spectra and light 
curves of SNe~Ia \citep{Hoeflich1996b,Nugent1997a}.

The outline of the paper is as follows: in Section~\ref{sec:models} we 
give a short summary of the models of \citet{Fink2010a} before briefly 
describing details of our radiative transfer simulations in Section~\ref{sec:rt}.
In Section~\ref{sec:oo} we present synthetic observables for the \citet{Fink2010a} 
models and compare them to the observed properties of SNe~Ia. The results 
of this comparison and implications for future work on sub-Chandrasekhar-mass 
double detonations are discussed in Section~\ref{sec:discussion}. 
Finally, we draw conclusions in Section~\ref{sec:conclusions}.

\section{Models} 
\label{sec:models}
\begin{figure*}
  \centering
  \plotone{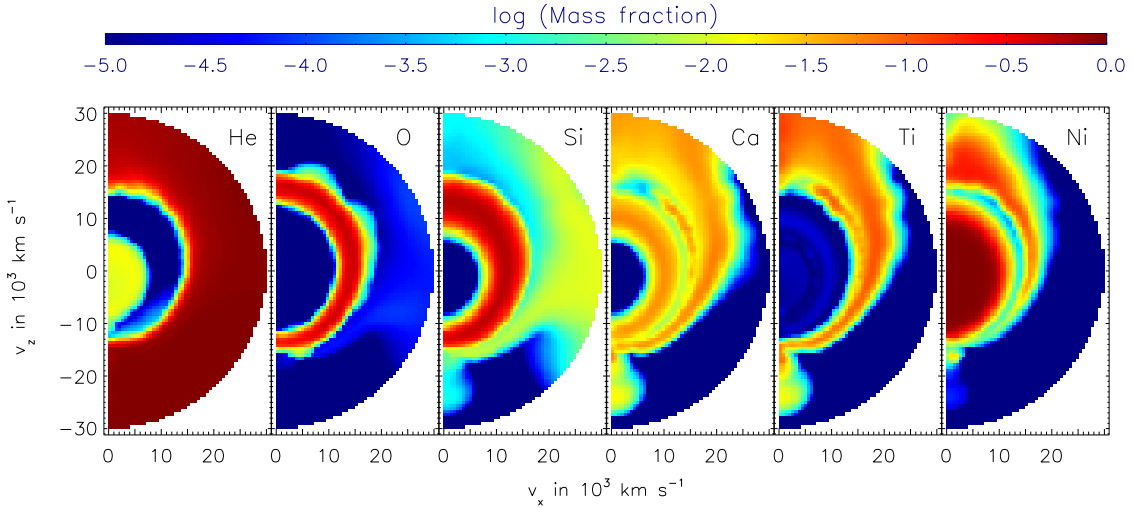}
  \caption{Final composition structure of selected species of Model 3 
    of \citet{Fink2010a}. The individual panels show the mass fractions 
    of He, O, Si, Ca, Ti and Ni (from left to right, respectively).
    The model is radially symmetric about the $z$-axis.}
  \label{fig:m03_composition}
\end{figure*}

Adopting the minimum helium shell masses required to initiate a helium 
detonation in the shell according to \citet{Bildsten2007a},
\citet{Fink2010a} investigated the double detonation scenario for six 
models representing a range of different C/O core masses (the model parameters 
are summarized in Table~\ref{tab:modelparas}). In all their models, they 
ignite an initial helium detonation in a single point at the base of 
the helium shell located on the positive $z$-axis (hereafter referred to 
as the ``north-pole'' of the WD). From there the helium detonation wave 
sweeps around the WD core until it converges at the south pole. At the 
same time a shock wave propagates into the core and converges at a 
point off-centre. Finding conditions that might be sufficient to initiate
a core detonation in a finite volume around this point, \citet{Fink2010a} 
then trigger a secondary core detonation at that point. This secondary 
detonation disrupts the entire WD and yields ejecta with a characteristic 
abundance distribution which is shown for Model 3 of \citet{Fink2010a} 
in Figure~\ref{fig:m03_composition}.

\begin{deluxetable*}{lcccccccc}
  \tabletypesize{\scriptsize}
  \tablecaption{Parameters of our model sequence.\label{tab:modelparas}}
  \tablehead{Model & \colhead{1} & \colhead{2} & \colhead{3} & \colhead{4} & \colhead{5} & \colhead{6} & \colhead{3c} & \colhead{3m}}

  \startdata
    $M_\mathrm{tot}$     & 0.936  & 1.004  & 1.080  & 1.164  & 1.293  & 1.3885 & 1.025 & 1.071\\
    \tableline
    $M_\mathrm{core}$    & 0.810  & 0.920  & 1.025  & 1.125  & 1.280  & 1.3850 & 1.025 & 1.025\\
    $M_\mathrm{core}(^{56}\mathrm{Ni})\tablenotemark{a}$ & $1.7 \times 10^{-1}$ & $3.4 \times 10^{-1}$ & $5.5 \times 10^{-1}$ & $7.8 \times 10^{-1}$ & $1.05$ & $1.10$ & $5.5 \times 10^{-1}$ & $5.6 \times 10^{-1}$ \\
    $M_\mathrm{core}(^{52}\mathrm{Fe})\tablenotemark{a}$ & $7.6 \times 10^{-3}$ & $9.9 \times 10^{-3}$ & $9.6 \times 10^{-3}$ & $7.9 \times 10^{-3}$ & $4.2 \times 10^{-3}$ & $1.7 \times 10^{-4}$ & $9.6 \times 10^{-3}$ & $9.4 \times 10^{-3}$\\
    $M_\mathrm{core}(^{48}\mathrm{Cr})\tablenotemark{a}$ & $3.9 \times 10^{-4}$ & $4.6 \times 10^{-4}$ & $4.5 \times 10^{-4}$ & $3.8 \times 10^{-4}$ & $2.1 \times 10^{-4}$ & $7.1 \times 10^{-5}$ & $4.5 \times 10^{-4}$ & $4.4 \times 10^{-4}$ \\
    $M_\mathrm{core}(^{44}\mathrm{Ti})$ & $7.2 \times 10^{-6}$ & $9.0 \times 10^{-6}$ & $1.1 \times 10^{-5}$ & $1.4 \times 10^{-5}$ & $1.4 \times 10^{-5}$ & $9.9 \times 10^{-6}$ & $1.1 \times 10^{-5}$ & $1.2 \times 10^{-5}$ \\
    $M_\mathrm{core}(^{40}\mathrm{Ca})$ & $2.0 \times 10^{-2}$ & $2.1 \times 10^{-2}$ & $1.8 \times 10^{-2}$ & $1.4 \times 10^{-2}$ & $6.9 \times 10^{-3}$ & $1.8 \times 10^{-3}$ & $1.8 \times 10^{-2}$ & $1.8 \times 10^{-2}$ \\
    $M_\mathrm{core}(^{36}\mathrm{Ar})$ & $2.2 \times 10^{-2}$ & $2.2 \times 10^{-2}$ & $1.9 \times 10^{-2}$ & $1.5 \times 10^{-2}$ & $6.7 \times 10^{-3}$ & $1.7 \times 10^{-3}$ & $1.9 \times 10^{-2}$ & $1.9 \times 10^{-2}$ \\
    $M_\mathrm{core}(^{32}\mathrm{S})$  & $1.3 \times 10^{-1}$ & $1.2 \times 10^{-1}$ & $1.0 \times 10^{-1}$ & $7.5 \times 10^{-2}$ & $3.3 \times 10^{-2}$ & $8.0 \times 10^{-3}$ & $1.0 \times 10^{-1}$ & $1.0 \times 10^{-1}$ \\
    $M_\mathrm{core}(^{28}\mathrm{Si})$ & $2.7 \times 10^{-1}$ & $2.5 \times 10^{-1}$ & $2.1 \times 10^{-1}$ & $1.4 \times 10^{-1}$ & $6.1 \times 10^{-2}$ & $1.5 \times 10^{-2}$ & $2.1 \times 10^{-1}$ & $2.0 \times 10^{-1}$ \\
    $M_\mathrm{core}(^{24}\mathrm{Mg})$ & $4.5 \times 10^{-2}$ & $3.5 \times 10^{-2}$ & $2.4 \times 10^{-2}$ & $1.1 \times 10^{-2}$ & $9.3 \times 10^{-3}$ & $4.3 \times 10^{-3}$ & $2.4 \times 10^{-2}$ & $2.2 \times 10^{-2}$ \\
    $M_\mathrm{core}(^{16}\mathrm{O})$  & $1.4 \times 10^{-1}$ & $1.1 \times 10^{-1}$ & $8.0 \times 10^{-2}$ & $4.2 \times 10^{-2}$ & $3.1 \times 10^{-2}$ & $1.2 \times 10^{-2}$ & $8.0 \times 10^{-2}$ & $7.7 \times 10^{-2}$ \\
    $M_\mathrm{core}(^{12}\mathrm{C})$  & $6.6 \times 10^{-3}$ & $4.4 \times 10^{-3}$ & $2.7 \times 10^{-3}$ & $8.8 \times 10^{-4}$ & $5.9 \times 10^{-3}$ & $7.4 \times 10^{-4}$ & $2.7 \times 10^{-3}$ & $3.3 \times 10^{-3}$ \\
    \tableline
    $M_\mathrm{sh}$      & 0.126  & 0.084  & 0.055  & 0.039  & 0.013  & 0.0035 & -- & 0.046\\
    $M_\mathrm{sh}(^{56}\mathrm{Ni})$\tablenotemark{a}   & $8.4 \times 10^{-4}$ & $1.1 \times 10^{-3}$ & $1.7 \times 10^{-3}$ & $4.4 \times 10^{-3}$ & $1.5 \times 10^{-3}$ & $5.7 \times 10^{-4}$ & -- & $1.1 \times 10^{-8}$\\
    $M_\mathrm{sh}(^{52}\mathrm{Fe})$\tablenotemark{a}   & $7.6 \times 10^{-3}$ & $7.0 \times 10^{-3}$ & $6.2 \times 10^{-3}$ & $3.5 \times 10^{-3}$ & $1.2 \times 10^{-3}$ & $2.0 \times 10^{-4}$ & -- & $6.1 \times 10^{-9}$\\
    $M_\mathrm{sh}(^{48}\mathrm{Cr})$\tablenotemark{a}   & $1.1 \times 10^{-2}$ & $7.8 \times 10^{-3}$ & $4.4 \times 10^{-3}$ & $2.2 \times 10^{-3}$ & $6.8 \times 10^{-4}$ & $1.5 \times 10^{-4}$ & -- & $7.9 \times 10^{-8}$\\
    $M_\mathrm{sh}(^{44}\mathrm{Ti})$   & $7.9 \times 10^{-3}$ & $5.4 \times 10^{-3}$ & $3.4 \times 10^{-3}$ & $1.8 \times 10^{-3}$ & $4.9 \times 10^{-4}$ & $6.2 \times 10^{-5}$ & -- & $4.2 \times 10^{-5}$ \\
    $M_\mathrm{sh}(^{40}\mathrm{Ca})$   & $4.7 \times 10^{-3}$ & $3.2 \times 10^{-3}$ & $2.2 \times 10^{-3}$ & $2.2 \times 10^{-3}$ & $6.8 \times 10^{-4}$ & $2.4 \times 10^{-4}$ & -- & $3.5 \times 10^{-3}$\\
    $M_\mathrm{sh}(^{36}\mathrm{Ar})$   & $5.0 \times 10^{-3}$ & $3.2 \times 10^{-3}$ & $2.0 \times 10^{-3}$ & $1.6 \times 10^{-3}$ & $5.2 \times 10^{-4}$ & $1.3 \times 10^{-4}$ & -- & $2.7 \times 10^{-2}$\\
    $M_\mathrm{sh}(^{32}\mathrm{S})$    & $2.2 \times 10^{-3}$ & $1.2 \times 10^{-3}$ & $7.8 \times 10^{-4}$ & $1.3 \times 10^{-3}$ & $4.3 \times 10^{-4}$ & $1.9 \times 10^{-4}$ & -- & $1.4 \times 10^{-2}$\\
    $M_\mathrm{sh}(^{28}\mathrm{Si})$   & $4.8 \times 10^{-4}$ & $2.5 \times 10^{-4}$ & $1.4 \times 10^{-4}$ & $4.7 \times 10^{-4}$ & $1.6 \times 10^{-4}$ & $1.3 \times 10^{-4}$ & -- & $5.9 \times 10^{-4}$\\
    $M_\mathrm{sh}(^{24}\mathrm{Mg})$   & $4.3 \times 10^{-5}$ & $2.3 \times 10^{-5}$ & $1.3 \times 10^{-5}$ & $1.0 \times 10^{-4}$ & $3.8 \times 10^{-5}$ & $3.6 \times 10^{-5}$ & -- & $3.0 \times 10^{-5}$\\
    $M_\mathrm{sh}(^{16}\mathrm{O})$    & $3.2 \times 10^{-6}$ & $1.7 \times 10^{-6}$ & $1.9 \times 10^{-6}$ & $6.0 \times 10^{-5}$ & $2.1 \times 10^{-5}$ & $2.6 \times 10^{-5}$ & -- & $1.0 \times 10^{-5}$\\
    $M_\mathrm{sh}(^{12}\mathrm{C})$    & $1.2 \times 10^{-3}$ & $5.3 \times 10^{-4}$ & $2.2 \times 10^{-4}$ & $7.9 \times 10^{-5}$ & $1.7 \times 10^{-5}$ & $1.6 \times 10^{-6}$ & -- & $3.2 \times 10^{-5}$\\
    $M_\mathrm{sh}(^{4}\mathrm{He})$    & $8.3 \times 10^{-2}$ & $5.3 \times 10^{-2}$ & $3.3 \times 10^{-2}$ & $2.0 \times 10^{-2}$ & $6.9 \times 10^{-3}$ & $1.7 \times 10^{-3}$ & -- & $2.7 \times 10^{-4}$\\
    \tableline

    $\Delta m_{15}(B)$ / mag    & 0.88  & 1.25  & 1.74  & 1.77  & 1.23  & 1.29  & 1.63  & 1.37\\
    $t_\mathrm{max}(B)$ / days  & 17.0  & 17.0  & 17.7  & 16.4  & 15.2  & 14.3  & 18.2  & 18.1\\
    $M_\mathrm{B,max}$ / mag      & -15.9 & -17.3 & -18.4 & -19.3 & -19.8 & -19.9 & -19.2 & -19.1\\
    $M_\mathrm{V,max}$ / mag      & -17.6 & -18.8 & -19.6 & -19.9 & -20.1 & -20.1 & -19.4 & -19.4\\
    $M_\mathrm{R,max}$ / mag      & -18.4 & -19.1 & -19.4 & -19.4 & -19.4 & -19.2 & -19.0 & -19.1\\
    $M_\mathrm{I,max}$ / mag      & -18.9 & -19.2 & -19.2 & -19.4 & -19.6 & -19.7 & -18.9 & -19.0\\
    $(U-B)_\mathrm{Bmax}$ / mag  & 0.58  & 0.41  & 0.50  & 0.19  & 0.08  & -0.26 & -0.04 & 0.35\\
    $(B-V)_\mathrm{Bmax}$ / mag  & 1.67  & 1.51  & 1.13  & 0.59  & 0.28  & 0.08  & 0.11  & 0.30\\
    $(V-R)_\mathrm{Bmax}$ / mag  & 0.74  & 0.36  & -0.13 & -0.46 & -0.67 & -0.82 & -0.30 & -0.28\\
    $(V-I)_\mathrm{Bmax}$ / mag  & 1.26  & 0.39  & -0.47 & -1.05 & -1.48 & -1.47 & -0.64 & -0.49\\
  \enddata
  
  \tablenotetext{a}{At maximum light $^{56}$Ni, $^{52}$Fe and $^{48}$Cr
    will have mostly decayed to $^{56}$Co, $^{52}$Cr and a mixture of 
    $^{48}$V and $^{48}$Ti, respectively.}
  \tablecomments{$M_\mathrm{tot}$,
    $M_\mathrm{core}$, and $M_\mathrm{sh}$ are the masses of the WD,
    the C/O core, and the helium shell, respectively.  All masses are given 
    in units of the solar mass. $\Delta m_{15}(B)$ and $t_\mathrm{max}(B)$ 
    refer to the decline parameter and the rise time to maximum light in 
    the angle-averaged $B$-band light curves, respectively. $M_\text{B,max}$, 
    $M_\text{V,max}$, $M_\text{R,max}$ and $M_\text{I,max}$ denote the angle-averaged
    peak magnitudes at the true peaks in the given bands. Colours are quoted at 
    time [$t_\mathrm{max}(B)$] of $B$-band maximum.}
\end{deluxetable*}

In the initial helium shell the burning does not reach nuclear statistical 
equilibrium due to the low densities. Thus it mainly produces iron-group 
elements lighter than $^{56}$Ni such as titanium, chromium and iron -- 
including some amount of the radioactive isotopes $\mathrm{^{48}Cr}$ and 
$\mathrm{^{52}Fe}$. Aside from a small mass of calcium, no significant 
amounts of intermediate-mass elements are produced in the shell. However, 
a large fraction of helium remains unburned. 

The core detonation yields both iron-group and intermediate-mass elements,
the relative amounts of which depend crucially on the core density and thus 
WD mass. Models with more massive WDs produce more iron-group material and 
less intermediate-mass elements (similar to the explosions of the naked 
sub-Chandrasekhar-mass WDs studied in \citealt{Sim2010a}). The most massive 
Model 6 of the sequence of \citet{Fink2010a} produces almost only iron-group 
material and hardly any intermediate-mass elements (the most important burning 
products in the shell and core of the different models are listed in 
Table~\ref{tab:modelparas}).

Due to the single-point ignition the models show strong ejecta asymmetries
which can be divided into two main categories. First, the helium shell 
contains more iron-group material on the northern hemisphere as discussed
by \citet{Fink2010a}. Second, the ignition point of the secondary core 
detonation is offset from the centre-of-mass of the model due to the 
off-centre convergence of the shock waves from the helium detonation.

\section{Radiative transfer simulations} 
\label{sec:rt}

To derive synthetic observables for the models, we performed radiative
transfer simulations with the time-dependent multi-dimensional Monte 
Carlo radiative transfer code {\sc artis} described by \citet{Kromer2009a}
and \citet{Sim2007a}. Since the models produce significant amounts of 
$\mathrm{^{48}Cr}$ and $\mathrm{^{52}Fe}$, we extended {\sc artis} to 
take into account the energy released by the decay sequences $^{52}\mbox{Fe}\rightarrow{^{52}\mbox{Mn}}\rightarrow{^{52}\mbox{Cr}}$
and 
$^{48}\mbox{Cr}\rightarrow{^{48}\mbox{V}}\rightarrow{^{48}\mbox{Ti}}$
in addition to the radioactive decays of $^{56}\mbox{Ni}\rightarrow{^{56}\mbox{Co}}$ 
and $^{56}\mbox{Co}\rightarrow{^{56}\mbox{Fe}}$ which form the primary 
energy source of SNe Ia (\citealt{Truran1967a}, \citealt{Colgate1969a}).
For other radioactive nuclei which are synthesized during the explosion 
in a non-negligible amount (e.g. $^{44}\mbox{Ti}$ in the shell) the life
times are much longer than those of the $^{56}$Ni decay-sequence. Thus 
they can be neglected at early times when the decays of $^{56}$Ni and 
$^{56}$Co power the light curves.

The total $\gamma$-ray energy $E_\mathrm{tot}$ that will be emitted from 
$t=0$ to $t\rightarrow\infty$ in the decay chains we consider, will be
\begin{align}
  E_{\mathrm{tot}} &=(E_{^{56}\mathrm{Ni}}+E_{^{56}\mathrm{Co}})\frac{M_{^{56}\mathrm{Ni}}}{m_{^{56}\mathrm{Ni}}}+ (E_{^{52}\mathrm{Fe}}+E_{^{52}\mathrm{Mn}})\frac{M_{^{52}\mathrm{Fe}}}{m_{^{52}\mathrm{Fe}}}\notag\\
 & \quad +(E_{^{48}\mathrm{Cr}}+E_{^{48}\mathrm{V}})\frac{M_{^{48}\mathrm{Cr}}}{m_{^{48}\mathrm{Cr}}}.
  \label{eq:decays}
\end{align}
Here $E_{\mathrm{^{56}Ni}}$, ($E_{^{56}\mathrm{Co}}$, $E_{^{52}\mathrm{Fe}}$, 
$E_{^{52}\mathrm{Mn}}$, $E_{^{48}\mathrm{Cr}}$, $E_{^{48}\mathrm{V}}$)
is the mean energy emitted per decay of $^{56}$Ni, ($^{56}$Co,
$^{52}$Fe, $^{52}$Mn, $^{48}$Cr, $^{48}$V). Similarly,
$M_{\mathrm{^{56}Ni}}$ ($M_{\mathrm{^{52}Fe}}$, $M_{\mathrm{^{48}Cr}}$
) is the initial mass of $^{56}$Ni ($^{52}\mbox{Fe}$, $^{48}\mbox{Cr}$)
synthesized in the explosion and $m_{\mathrm{^{56}Ni}}$ ($m_{\mathrm{^{52}Fe}}$,
$m_{\mathrm{^{48}Cr}}$) the mass of the $^{56}$Ni ($^{52}\mbox{Fe}$,
$^{48}\mbox{Cr}$) atom.

Following \cite{Lucy2005a}, this energy is quantized into
$\mathcal{N}=E_{\mathrm{tot}}/\epsilon_{0}$ identical indivisible energy 
``pellets'' of initial co-moving frame (cmf) energy $\epsilon_{0}$. The 
pellets are first assigned to one of the decay sequences in proportion to 
the amount of energy deposited in the different decay sequences (terms 
on the right-hand-side of Equation~\ref{eq:decays}). Then they are distributed 
on the grid according to the initial distribution of $^{56}$Ni, $^{52}$Fe 
and $^{48}$Cr, as appropriate, and follow the homologous expansion until 
they decay. Decay times are sampled in a two step process. If the pellet 
was assigned to the $^{56}$Ni chain, for example, we first choose whether
it belongs to a decay of the parent nucleus $^{56}$Ni or the daughter 
nucleus $^{56}$Co by sampling the probabilities 
$E_{\mathrm{^{56}Ni}}/\left(E_{\mathrm{^{56}Ni}}+E_{\mathrm{^{56}Co}}\right)$
and 
$E_{\mathrm{^{56}Co}}/\left(E_{\mathrm{^{56}Ni}}+E_{\mathrm{^{56}Co}}\right)$,
respectively. Finally an appropriate decay time is sampled 
\begin{equation}
t_{\mathrm{decay}}\left(^{56}\mathrm{Ni}\right)=-\tau\left(^{56}\mathrm{Ni}\right)\log\left(z_{1}\right)
\end{equation}
\begin{equation}
t_{\mathrm{decay}}\left(^{56}\mathrm{Co}\right)=-\tau\left(^{56}\mathrm{Ni}\right)\log\left(z_{1}\right)-\tau\left(^{56}\mathrm{Co}\right)\log\left(z_{2}\right)
\end{equation}
from the mean life times of the $^{56}$Ni [$\tau\left(^{56}\mathrm{Ni}\right)=8.80\,\mbox{d}$]
and $^{56}$Co [$\tau\left(^{56}\mathrm{Co}\right)=113.7\,\mbox{d}$] nuclei. $z_i$ are
random numbers between 0 and 1. The $^{52}\mbox{Fe}$ [$\tau\left(^{52}\mathrm{Fe}\right)=0.4974\,\mbox{d}$,
$\tau\left(^{52}\mathrm{Mn}\right)=0.02114\,\mbox{d}$] and $^{48}\mbox{Cr}$
[$\tau\left(^{48}\mathrm{Cr}\right)=1.296\,\mbox{d}$, $\tau\left(^{48}\mathrm{V}\right)=23.04\,\mbox{d}$]
chains are treated in the same way.

Upon decay, a pellet transforms to a single $\gamma$-packet representing 
a bundle of monochromatic-radiation of cmf energy $\epsilon_0$. The cmf 
photon energy of the $\gamma$-packets is randomly sampled from the 
relative probabilities of the $\gamma$-lines in the appropriate decay 
of the selected decay sequence -- including annihilation lines due to 
positron emission. Following \citet{Lucy2005a}, we assume that positrons 
released by radioactive decays annihilate in situ, giving rise to the 
emission of two 511 keV $\gamma$-ray photons. In doing so, we neglect the 
kinetic energy released by stopping the positrons, any positron escape and 
possible positronium formation which gives rise to the emission of continuum  
photons (see discussion by \citealt{Milne2004a}). Thus, our prediction of the 
511 keV line flux should be considered as an upper limit. The $\gamma$-packets 
are then propagated through the ejecta as described by \cite{Kromer2009a}. 

The $\gamma$-line data is taken from Table~1 of \citet{Ambwani1988a} for the 
$^{56}$Ni decay-sequence and from \citet{Burrows2006a} for the $^{48}$Cr 
decay-sequence, respectively. Owing to the comparatively short life times 
of $^{52}$Fe and $^{52}$Mn, these nuclei have mostly already decayed to 
their daughter nuclei when we start the radiative transfer simulation at 
$\sim 1$ day. Since the ejecta at these early times are almost opaque to 
$\gamma$-rays, the $\gamma$-packets released by the $^{52}$Fe decay chain 
will be thermalized rapidly. Therefore we do not follow the propagation 
of the $\gamma$-packets released by the $^{52}$Fe decay chain, but 
immediately convert their energy to thermal kinetic energy  ($k$-packets 
in the framework of {\sc artis}).

The input models (composition, density, velocities) for the radiative
transfer simulations were derived using the tracer particles from
the hydrodynamics simulations (see \citealt{Fink2010a} for details
on the hydro setup and use of tracer particles for nucleosynthesis).
Since the tracer particles are Lagrangian, we use an approach similar
to the reconstruction of the density field in smooth particle 
hydrodynamics (SPH) simulations to construct the input model from
the tracers. For the models described here, the density field was 
obtained using the SPH method described by \citet{Dolag2009a} 
(specifically, their equations 1 -- 3) adopting $N = 32$ for the SPH 
smoothing-length normalisation factor. 
For the centre ({\boldmath ${x_i}$}) of each grid cell ($i$) in the 
model, the mass fractions ($X_{Z,i}$) of the elements considered 
($Z = 1$ to 30) were reconstructed using
\begin{equation}
X_{Z,i} = \rho_{i} \sum_j W(| {\mbox{\boldmath $x_{i} - x_{j}$}} |, h_{j})X_{Z,j},
\end{equation}
where $\rho_{i}$ is the reconstructed mass density, $X_{Z,j}$ is the 
mass-fraction of element $Z$ for tracer particle $j$ (which lies at 
position {\boldmath $x_{j}$}), $h_{j}$ is the SPH particle smoothing 
length and $W(x,h)$ the SPH kernel function (defined via equations 3 
and 1 of \citealt{Dolag2009a}, respectively). The mass fractions of 
the important radioactive isotopes ($^{56}$Ni, $^{56}$Co, $^{52}$Fe 
and $^{48}$Cr) were reconstructed from the tracer particle yields 
in exactly the same manner. The reconstruction was performed on an 
$80\times160$ $(r,z)$-grid using the final state of the tracer 
particles at the end of the simulations (which were run up to the 
phase of homologous expansion). 

For the radiative transfer simulation this 2D model is re-mapped to a 
3D Cartesian grid of size $100^{3}$ which co-expands with the ejecta. 
We then follow the radiative transfer from 2 to 120 days after the 
explosion, discretized into 111 logarithmic time steps each of duration
$\Delta\ln\left(t\right)=0.037$. Using our detailed ionization treatment 
and the cd23\_gf-5 atomic data set \citep{Kromer2009a}, we simulated 
the propagation of $2\times10^{7}$ packets. To speed up the initial 
phase of the calculation we made use of our initial grey approximation 
for the first 30 time steps (adopting a parameterised grey opacity 
for the highly optically thick cells in the centre of our simulation
volume). The first ten time steps were treated in local thermodynamic
equilibrium (LTE).

\section{Synthetic observables}
\label{sec:oo}
In this Section we present synthetic observables for the models of 
\citet{Fink2010a}. First we consider angle-averaged ultraviolet, optical 
and infrared light curves and explore the diversity within this set of
models and compare it to that of observed SNe Ia (Section~\ref{sub:oo_lcs}).
Then we investigate the colour evolution and spectra of these models
(Sections~\ref{sub:oo_colours} and \ref{sub:oo_spectra}). 
In Section~\ref{sub:oo_los} we study the effects of the asymmetric ejecta 
composition. Finally, in Section~\ref{sub:oo_gamma}, we briefly discuss 
the $\gamma$-ray emission from these models.

\subsection{Broad-band light curves}
\label{sub:oo_lcs}
\begin{figure*}
  \centering
  \plotone{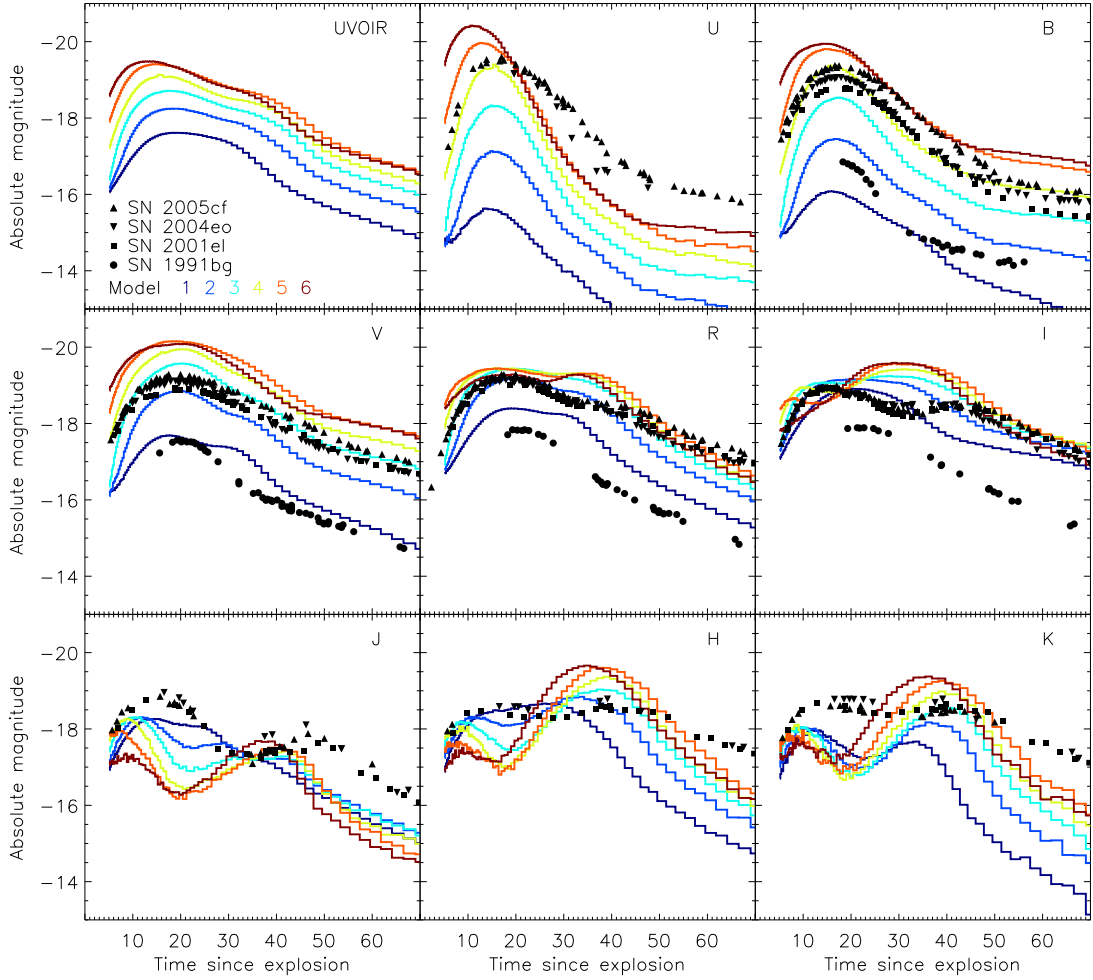}
  \caption{Angle-averaged $UVOIR$ bolometric and $U$,$B$,$V$,$R$,$I$,$J$,$H$,$K$ 
    band-limited light curves for the model sequence of \citet{Fink2010a} 
    as indicated by the colour coding. For comparison photometrical data
    of the spectroscopically normal SNe 2005cf (triangles; \citealt{Pastorello2007a}),
    2004eo (upside-down triangles; \citealt{Pastorello2007b}) and 2001el (squares;
    \citealt{Krisciunas2003a}) as well as for the subluminous SN 1991bg (circles; 
    \citealt{Filippenko1992b}, \citealt{Leibundgut1993a}) are shown.}
  \label{fig:lightcurves_all}
\end{figure*}

In Figure~\ref{fig:lightcurves_all} we show the angle-averaged 
ultraviolet-optical-infrared ($UVOIR$) bolometric and $U$, $B$, $V$, 
$R$, $I$, $J$, $H$, and $K$ band-limited light curves for the model 
sequence of \citet{Fink2010a} as obtained from our radiative transfer 
simulations. For comparison to observations we included photometric 
data of SNe~2005cf \citep{Pastorello2007a}, 2004eo \citep{Pastorello2007b},
2001el \citep{Krisciunas2003a} and 1991bg \citep{Filippenko1992b,Leibundgut1993a} 
in the figure. While SN~1991bg, the prototypical event of the subluminous 
1991bg-like objects, marks the faint end of observed SNe Ia, SNe~2005cf, 
2004eo and 2001el are representative of the spectroscopically normal objects.

\begin{figure*}
  \centering
  \plotone{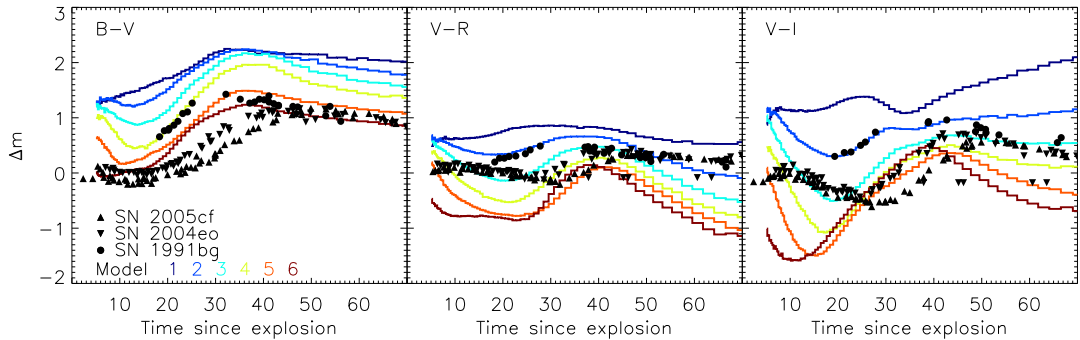}
  \caption{Angle-averaged colour curves of our models as indicated by the labels. 
    The different panels correspond to $B-V$, $V-R$ and $V-I$ colour (from left to
    right). For comparison, the colours of the spectroscopically normal SNe 2005cf 
    (triangles; \citealt{Pastorello2007a}) and 2004eo (upside-down triangles; 
    \citealt{Pastorello2007b}) as well as for the the subluminous SN 1991bg (circles; 
    \citealt{Filippenko1992b}, \citealt{Leibundgut1993a}) are shown.}
  \label{fig:colours_all}
\end{figure*}

Since our models form a sequence of increasing $^{56}\mathrm{Ni}$-mass, 
the peak brightness of the synthetic light curves increases from Model 1 
to Model 6: for $B$-band maximum, for example, we find a range from 
$-15.9\,\mathrm{mag}$ for Model 1 to $-19.9\,\mathrm{mag}$ for Model 6
(values for the intermediate models and other bands are given in
Table~\ref{tab:modelparas}). This covers almost the full range of
peak brightnesses observed among SNe Ia \citep[e.g.][]{Hicken2009a}, excluding
only the very brightest events like SN~2003fg \citep{Howell2006a}
which have been suggested to derive from super-Chandrasekhar-mass 
progenitor systems (see however \citealt{Hillebrandt2007a,Sim2007b}).

Despite their lower total mass, and thus lower overall opacity, our model 
light curves do not show particularly fast evolution compared to 
Chandrasekhar-mass models. In contrast, \citet{Hoeflich1996b} reported
a fast initial rise and broad peaks for their helium detonation models. 
For the rise times from explosion to $B$-band maximum [$t_\mathrm{max}(B)$] 
we find values between 17.0 and 17.7 days for our least massive models 
(1 to 3, cf. Table~\ref{tab:modelparas}). This is in good agreement with 
observational findings (e.g. \citealt{Hayden2010a} find rise times to 
$B$-band maximum between 13 and 23 days with an average value of $17.38 
\pm 0.17\,\mathrm{days}$). In contrast, the more massive Models (4 to 6)
show shorter $B$-band rise times with increasing mass (see Table~\ref{tab:modelparas}):
for Model 6, $t_\mathrm{max}(B)$ is only 14.3 days. The peak time is 
mainly set by the diffusion time for photons to leak out from the 
$^{56}$Ni-rich inner core. Owing to the larger densities 
in the more massive models, nuclear burning produces $^{56}$Ni out to 
much higher velocities than in the lower mass models (compare Figure~3
of \citealt{Fink2010a}). Moreover, the helium shell masses in our models 
decrease for the heavier WDs. Thus, the mass (and therefore opacity) on top 
of the $^{56}$Ni-rich inner core decreases with increasing WD mass and 
photons start to escape earlier for the more massive models making their 
light curves peak faster (compare also the rise times to bolometric 
maximum in \citealt{Fink2010a}).
In addition, we note that the radioactive nuclides which are produced in
the helium shell burning have some influence on the initial rise phase
of the light curves (see Figure~8 of \citealt{Fink2010a}). 

In the post-maximum decline phase, characterized by the decline 
parameter $\Delta m_{15}(B)$ which gives the change in $B$-band 
magnitude between maximum and 15 days thereafter, our models show 
a peculiar characteristic. Observationally, brighter SNe Ia show
a trend of slower declining light curves (e.g. \citealt{Phillips1993a,Hicken2009a}).
In contrast, the model light curves decline faster along our 
sequence from Model 1 to 4 despite increasing brightness 
(cf. Table~\ref{tab:modelparas}).

Specifically, we find $\Delta m_{15}(B) \sim 0.88$ for Model 1, which 
according to peak brightness would be classified as a subluminous explosion. 
However, observationally these events are characterized by a fast decline 
[$\Delta m_{15}(B) \sim 1.9$, e.g. \citealt{Garnavich2004a,Taubenberger2008a}]. 
In contrast, Model 4, which has a $B$-band peak magnitude typical for a 
spectroscopically normal SN Ia, yields $\Delta m_{15}(B) \sim 1.77$ 
which is much faster than typically observed [$\Delta m_{15}(B) 
\sim 1$, e.g. \citealt{Hicken2009a}]. 
Although $\Delta m_{15}(B)$ is lower for the brighter Models 5 and 6,
they still decline too fast compared to observed objects of corresponding
brightness. A similar trend for more rapidly declining light curves 
from more massive models is also seen in the bolometric light 
curves of Model 2 to 6 (cf. Table~4 of \citealt{Fink2010a}).

In contrast to earlier work by \citet{Hoeflich1996b}, who 
studied sub-Chandrasekhar-mass models with significantly more massive 
shells, we do not find particularly fast evolution in the post-maximum 
decline of our light curves compared to those obtained for other types
of models. Applying our radiative transfer code {\sc artis} for example
to the well-known W7 model \citep{Nomoto1984a,Thielemann1986a}, which is 
widely regarded as a good standard for SNe Ia, we find $\Delta m_{15}(B) 
\sim 1.6$ when using the same atomic data set as adopted here. This 
value is comparable to our fastest declining sub-Chandrasekhar-mass 
model and also too fast compared to normal SNe~Ia.

In addition, we note that $\Delta m_{15}(B)$ is also quite sensitive 
to the details of the radiative transfer treatment. Using an atomic 
data set with $8\times10^6$ lines (the big\_gf-4 data set of \citealt{Kromer2009a}), 
for example, yields $\Delta m_{15}(B)\sim1.75$ for W7. In contrast,
a simulation using the atomic data set adopted in this study but
applying a pure LTE treatment for the excitation/ionization state
of the plasma yields $\Delta m_{15}(B)\sim1.95$ for W7. This shows 
that systematic uncertainties in the radiative transfer treatment can 
affect $\Delta m_{15}(B)$ by several tenths of a magnitude (see also 
the comparison of different radiative transfer codes in Figure~7 of 
\citealt{Kromer2009a}). Thus, we argue that there is no evidence
that our sub-Chandrasekhar-mass models fade too fast compared to 
other explosion models.

\subsection{Colour evolution}
\label{sub:oo_colours}
The most striking difference between our light curves and those of the 
comparison objects in Figure~\ref{fig:lightcurves_all} concerns their 
colour. To highlight this we show the angle-averaged time-evolution of 
the $B-V$, $V-R$ and $V-I$ colours for all our models in Figure~\ref{fig:colours_all} 
and compare them again to our fiducial SNe 2005cf, 2004eo, 2001el and 1991bg. 

In $B-V$ all our models show positive colour indices for the whole 
simulation period and a red peak at $\sim 40$ days after explosion. 
Contrary to observed SNe Ia \citep{Lira1995a}, however, we find no 
convergence of the different models at epochs after the red peak. 
Instead, our models at all times form a sequence of increasingly redder 
$B-V$ colour towards the fainter explosions. With the exception of Model 
6, all our models are generally too red compared to spectroscopically 
normal SNe~Ia. At maximum light, for example, we find $(B-V)$-values 
of 1.67 and 0.28 for Model 1 and Model 5, respectively (values for the 
other models are given in Table~\ref{tab:modelparas}). In contrast, 
spectroscopically normal SNe Ia are characterized by $(B-V)_\mathrm{max} 
\sim 0.0\,\mathrm{mag}$. 

Subluminous 1991bg-like objects show a redder $B-V$ colour before the 
red peak, reaching $B-V\sim 0.4\dots0.7\,\mathrm{mag}$ at $B$-band 
maximum \citep{Taubenberger2008a}. Although Model 4 can
reproduce this, it is not a good fit to 1991bg-like objects, since it 
is considerably too bright (compare Figure~\ref{fig:lightcurves_all}).
Our subluminous Models 1 and 2, on the other hand, are significantly redder
than observed ($B-V$ colours of 1.67 and 1.51 at maximum, respectively).
In contrast, \citet{Hoeflich1996b} and \citet{Hoeflich1996c} found too 
blue colours at maximum light for their subluminous sub-Chandrasekhar-mass 
models compared to 1991bg-like objects. 

\begin{figure*}
  \centering
  \plotone{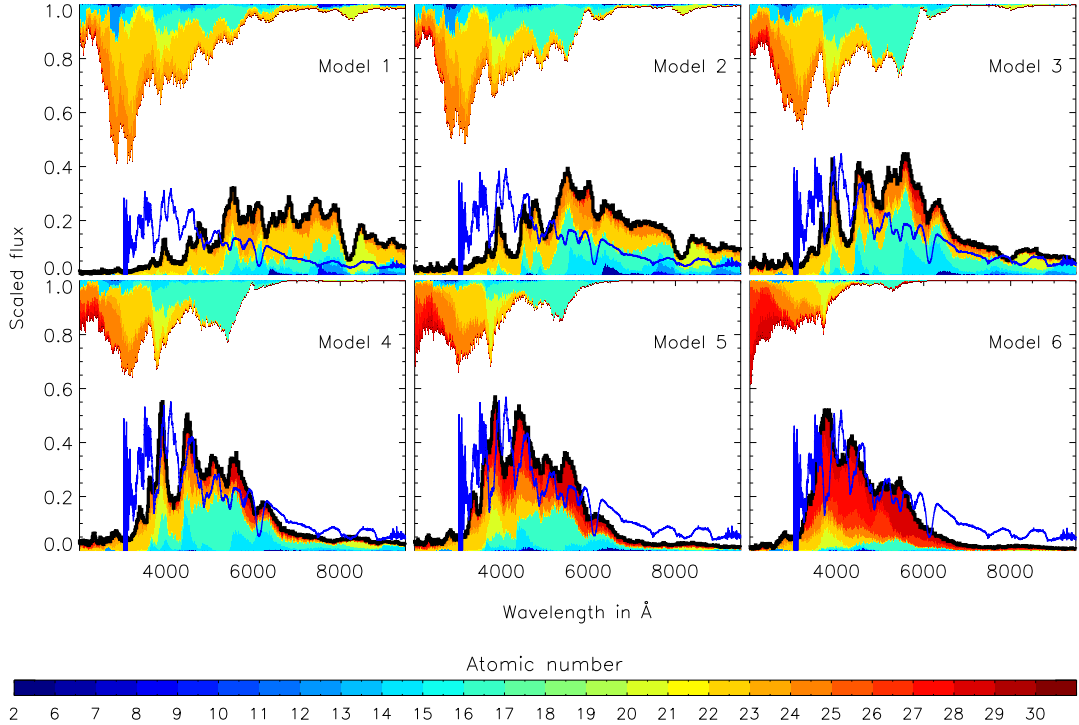}
  \caption{Angle-averaged (thick black line) spectra at three days before 
    $B$-band maximum for all six of our models as indicated by the labels. 
    For comparison the blue line shows the de-redshifted and de-reddened 
    spectrum of SN~2004eo \citep{Pastorello2007b} at the corresponding 
    epoch. This was scaled such that its maximum matches the maximum of 
    the model spectrum. The colour coding indicates the element(s) 
    responsible for both 
    bound-bound emission and absorption of quanta in the Monte Carlo 
    simulation. The region below the synthetic spectrum is colour coded
    to indicate the fraction of escaping quanta in each wavelength bin
    which last interacted with a particular element (the associated atomic
    numbers are illustrated in the colour bar). Similarly, the coloured
    regions above the spectra indicate which elements were last responsible
    for removing quanta from the wavelength bin (either by absorption
    or scattering/fluorescence).}
  \label{fig:maxspectra_all}
\end{figure*}

The origin of our red colours traces back to the extended layer of titanium 
and chromium which is present in the helium shell ejecta of our models. This 
will be discussed in more detail in Section~\ref{sub:oo_spectra}.

While the trend of increasingly redder colours for fainter models persists 
in the $V-R$ index, our models are not systematically too red here. Model 2 
and 3 populate about the right range for 1991bg-like and spectroscopically
normal SNe Ia, respectively. However, the details of their $V-R$ evolution,
especially the initial decline present in all the models, do not match the
observations.

A similar behaviour is found for the $V-I$ colour where again Model 2 and 
3 lie closest to the observed colours of 1991bg-like and spectroscopically 
normal SNe Ia, respectively. However, again the agreement is imperfect.
For Model 2 this is most obvious at the latest epochs, where 1991bg-like 
objects start to become bluer while the model colour maintains a redward 
evolution. Moreover, the model shows a secondary blue minimum, due to the 
post-maximum plateau of the $V$-band light curve which is not observed in 
1991bg-like objects. In Model 3 the initial decline and the rise to the 
red peak are significantly different from the observed behaviour of 
spectroscopically normal SNe Ia.

\subsection{Spectra}
\label{sub:oo_spectra}
The properties of our model light curves can be understood from
consideration of the synthetic spectra. Figure~\ref{fig:maxspectra_all}
shows the angle-averaged spectrum of all our models at three days before
$B$-band maximum. For comparison, we also show the spectrum of SN~2004eo 
\citep{Pastorello2007b} at the corresponding epoch. The colour coding 
below the synthetic spectrum indicates the fraction of escaping quanta 
in each wavelength bin which were last emitted in bound-bound transitions 
of a particular element. Similarly, the coloured regions above the spectra 
indicate which elements were last responsible for removing quanta from a 
wavelength bin by bound-bound absorption. This coding allows us to both 
identify individual spectral features and track the effect of fluorescence 
on the spectrum formation directly. The contributions of bound-free and 
free-free emissions to the total flux are so small that they are not 
discernable in the figure.

From this it is immediately obvious that the colours of our models 
are due to fluorescence in titanium and chromium lines. Having a 
wealth of strong lines in the UV and blue, even small amounts of these 
elements effectively block radiation in the UV and blue and redistribute 
it to redder wavelengths where the optical depths are smaller and the 
radiation can escape. Compared to other explosion models, this effect 
is particularly strong in the \citet{Fink2010a} models, since they 
produce relatively large amounts of titanium and chromium in the outer 
layers during the initial helium detonation (cf. Table~\ref{tab:modelparas} 
and Figure~\ref{fig:m03_composition}; typical titanium and chromium yields 
for other explosion models are on the order of the yields from the core 
detonation). This also explains the trend for redder colours in the fainter 
models: according to Table~\ref{tab:modelparas}, the production of titanium 
and chromium in the shell increases continuously from Model 6 to Model 1.

Since this titanium and chromium layer is located at higher velocities 
than most of the intermediate mass elements (cf. Figure~\ref{fig:m03_composition}),
redistribution by titanium and chromium also dilutes the absorption 
features of intermediate-mass elements like silicon and sulphur by 
reprocessing flux into the relevant wavelength regions. This can be 
clearly seen in Figure~\ref{fig:maxspectra_all}. Although Models 1 and 
2 produce the largest amounts of silicon of all our models (0.27 and 
0.25 solar masses in the core, respectively), they show only a (very) 
weak Si\,{\sc ii} $\lambda 6355$ feature since this wavelength region 
is strongly polluted by flux redistribution from titanium and chromium. 
The only feature of intermediate-mass elements which is clearly visible 
in these models is the Ca\,{\sc ii} near-infrared (NIR) triplet at $\lambda\lambda 
8498,8542,8662$. This feature remains unburied since calcium exists co-spatially 
with titanium and chromium in the outer shell (Figure~\ref{fig:m03_composition}).

Of all our models, Model 3 [$M_\mathrm{core}(^{28}\mathrm{Si})=0.21\,M_\odot$] 
produces the strongest Si\,{\sc ii} $\lambda 6355$ feature. For more 
massive models the silicon yields from the core detonation drop 
dramatically since, owing to the higher densities, a larger fraction 
of the core is burned to iron-group elements (see Table~\ref{tab:modelparas} 
and Figure~3 of \citealt{Fink2010a}). As expected from studies of pure 
detonations of Chandrasekhar-mass WDs \citep{Arnett1971a}, the extreme 
case of Model 6 (almost the Chandrasekhar mass) produces no significant 
amounts of intermediate-mass elements. Thus the spectrum is totally 
dominated by iron-group elements. Since it shows no indication of any 
Si\,{\sc ii} $\lambda 6355$ feature, this model would not be classified 
as a SN~Ia.

Since none of our models give a good match to the colours and line strengths
of observed SNe~Ia, we will not discuss line velocities in detail. We note, 
however, that along the sequence from Model 1 to 5 there is a trend for 
higher line velocities of intermediate mass elements. This is obvious in the 
Si\,{\sc ii} $\lambda 6355$ feature of our models which is apparent in 
Figure~\ref{fig:maxspectra_all}. Along the model sequence (1 to 5), this line 
moves to shorter wavelengths compared to the observed Si\,{\sc ii} absorption 
feature of SN~2004eo. This arises because the inner boundary of the region 
rich in intermediate-mass elements moves to higher velocities with increasing 
mass due to the more complete burning in the inner regions (compare Figure~3 
of \citealt{Fink2010a}). 

Finally, we note that none of our model spectra show any indication of 
helium lines, despite our models having up to $\sim0.08\,M_\odot$ of 
helium in their outer layers (cf. Table~\ref{tab:modelparas}). Since 
helium has rather highly excited levels, this might simply be a 
consequence of our approximate treatment of the plasma state which 
neglects non-thermal excitation and ionization. We note, however, that 
despite using the {\sc phoenix} code which does include a treatment of 
non-thermal processes, \citet{Nugent1997a} also found no strong evidence 
of helium lines for models with even more ($0.2\,M_\odot$) helium in the 
outer shell.

\subsection{Line-of-sight dependence}
\label{sub:oo_los}
As discussed in Section~\ref{sec:models}, our models show strong ejecta 
asymmetries due to their ignition in a single-point at the north pole
of the WD (cf. Figure~\ref{fig:m03_composition}). These asymmetries are 
expected to have some influence on the observables along different 
lines-of-sight. Since they are characteristically the same for all our 
models, we first use Model 3 as an example to give a detailed discussion 
of line-of-sight dependent spectra and light curves of this model 
(Sections~\ref{sub:oo_los_spectra} and \ref{sub:oo_los_lightcurves}). 
In Section~\ref{sub:oo_los_all} we then present the variations in peak 
magnitudes and colours due to line-of-sight effects for all our models.

\subsubsection{Spectra}
\label{sub:oo_los_spectra}

\begin{figure}
  \centering
  \plotone{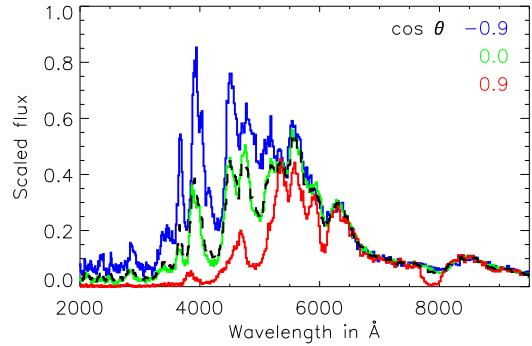}
  \caption{Line-of-sight dependent maximum-light spectra for Model 3.
    To indicate the maximal effect spectra seen pole-on are plotted 
    and compared to a spectrum seen equator on. The corresponding 
    lines are identified by the colour coding. For comparison the 
    angle-averaged spectrum is shown as the dashed black line.}
  \label{fig:m03_losspectra}
\end{figure}

To obtain line-of-sight dependent observables we bin the escaping photons
into a grid of ten equal solid-angle bins in $\mu=\cos\theta$ with $\theta$ 
being the angle between the line-of-sight and the $z$-axis of the model. 
In Figure~\ref{fig:m03_losspectra} we show the maximum-light spectra of 
Model 3 for three different directions, corresponding to lines-of-sight
close to the southern polar axis ($\mu=-0.9$), the northern polar axis ($\mu=
-0.9$) and equator-on (average of the $\mu=-0.1$ and 0.1 bins). While the 
equator-on spectrum looks similar to the angle-averaged spectrum, the 
spectra seen from the polar directions are significantly different from 
each other as well as from the angle-averaged spectrum. For $\mu=0.9$,
an observer is looking through the extended layer of iron-group elements 
in the outer shell of the northern-hemisphere (cf. Figure~\ref{fig:m03_composition}).
This blocks almost all the flux in the UV and blue wavelength range and
the redistribution of this flux leads to a very red spectrum. In contrast, 
this layer of iron-group elements in the outer shell is far less extended 
on the southern hemisphere, where the shell burning is less efficient. 
Thus, an observer with $\mu=-0.9$ sees a bluer spectrum than equator-on. 
Equator-on the extension of the layer of iron-group elements is somewhere
between these two extremes and the spectrum resembles the angle-averaged 
case.

\begin{figure*}
  \centering
  \plotone{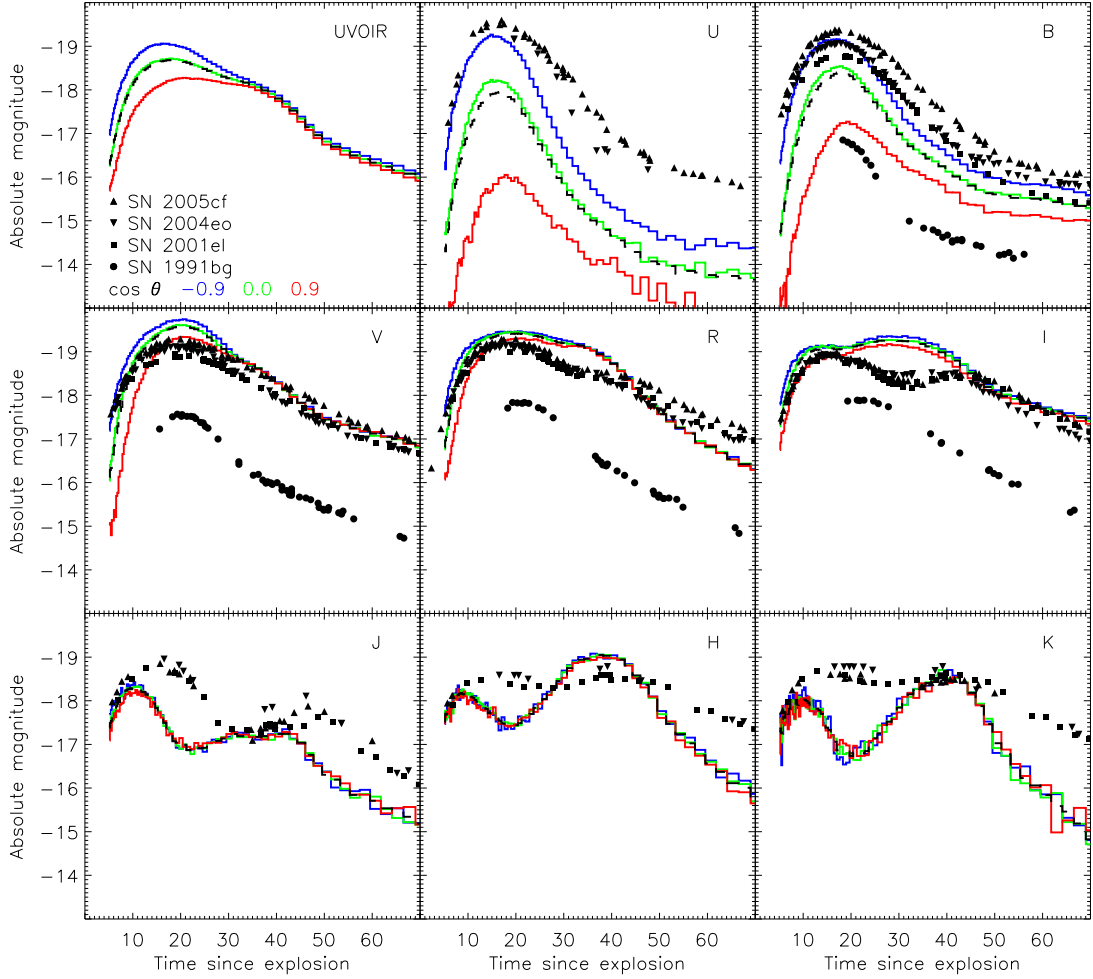}
  \caption{Selected line-of-sight dependent light curves of Model 3 as 
    indicated by the colour coding. For comparison angle-averaged
    light curves (black dashed) and photometrical data of our fiducial
    SNe 2005cf, 2004eo, 2001el and 1991bg (different symbols) are shown.}
  \label{fig:m03_lightcurves}
\end{figure*}

A secondary effect results from the off-centre ignition of the core. Since 
the core has been compressed less strongly on the northern hemisphere, the 
core detonation yields more intermediate-mass elements on the northern 
than on the southern hemisphere (cf. Figure~\ref{fig:m03_composition} and 
the Discussion in Section~4.3 of \citealt{Fink2010a}). This leads to 
stronger features of intermediate-mass elements in the spectrum seen along 
the northern polar-axis and becomes particularly visible in the strength 
of the Si\,{\sc ii} $6355\,\text{\AA}$ feature and the Ca\,{\sc ii} NIR-triplet 
which become weaker from $\mu=0.9 \rightarrow -0.9$. 

\newpage
\subsubsection{Lightcurves}
\label{sub:oo_los_lightcurves}
 
Figure~\ref{fig:m03_lightcurves} shows band-limited synthetic light 
curves from our radiative transfer simulations for Model 3 as seen 
equator-on and from the two polar directions. As already noted for 
the maximum-light spectra, our model shows a trend for increasingly 
red colours from $\mu=-0.9 \rightarrow 0.9$. Similarly, we find a 
clear dependence of the light curve rise and decline times on the
line-of-sight. While the rise times increase along the sequence 
$\mu=-0.9 \rightarrow 0.9$ (for $B$ band, for example, we find rise 
times between 16.8 and 19.7 days for $\mu=-0.9$ and $\mu=0.9$, 
respectively), the light curve declines more slowly [$\Delta m_{15}(B)=
1.91$ and $\Delta m_{15}(B)=1.29$ for $\mu=-0.9$ and $\mu=0.9$, 
respectively]. Moreover, we find a clear trend for a weaker dependence 
of the light curves on the line-of-sight at lower photon energies. 
Thus, while $U$ and $B$ band show a variation of $\sim3$ and $\sim2$ 
magnitudes at maximum light, respectively, the variation in $V$ band 
is already less than half a magnitude and in the NIR bands there is 
virtually no viewing-angle effect.

All these effects are due to the asymmetry of the layer of iron-group 
elements in the outer shell. Since this layer is more extended on the 
northern hemisphere it causes additional line blocking and thus enhanced 
fluorescence and photon trapping for inclinations close to $\mu=1$. This 
explains the redder colours as well as the increasing rise and decreasing 
decline times for $\mu=-0.9 \rightarrow 0.9$. For lower photon energies, 
the asymmetry of the outer shell is less important, since the optical 
depths are smaller and photons typically escape from deeper layers of 
the ejecta. In the NIR bands the entire ejecta contribute to the emitted 
photons and the viewing-angle dependence of the light curves is very small.

\begin{figure*}
  \centering
  \plotone{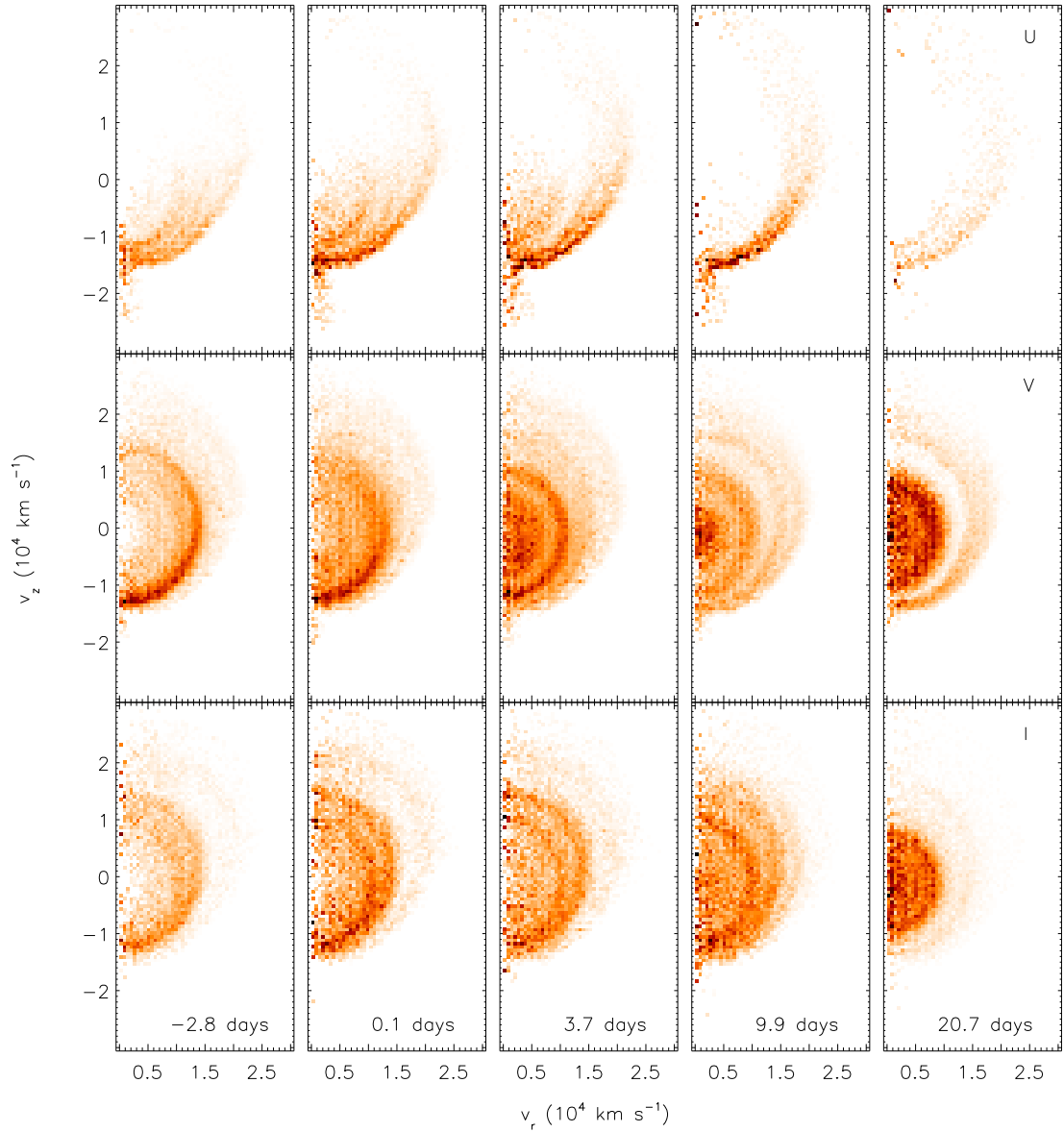}
  \caption{Region of last emission for selected bands ($U$, $V$, $I$
    from top to bottom) and different times (from left to right). The epochs
    indicated in the bottom panels are given in days with respect to $B$-band
    maximum. Dark regions contribute most to the flux escaping in the band. 
    The model is symmetric under rotation about the $z$-axis.}
  \label{fig:m03_lastscat_snapshots}
\end{figure*}

This is illustrated in Figure~\ref{fig:m03_lastscat_snapshots} which shows 
where photons of selected bands were last emitted before they escaped from 
the supernova ejecta at different times. While $U$-band photons leak out 
predominantly from the regions on the southern hemisphere, where the layer
of iron-group elements is least extended, $I$-band photons show no strong
preference for a particular region. In fact, even before maximum light
the whole ejecta contribute to $I$-band emission. $V$-band photons, in 
contrast, show some preference for leaking out from the southern hemisphere
before and immediately after maximum light. From about 10 days after
maximum light this preference disappears. This is directly reflected
in the $V$-band light curve in Figure~\ref{fig:m03_lightcurves} which 
becomes viewing-angle independent at about that time.

Finally, we note that the strong line-of-sight dependence of our synthetic
observables poses an additional problem for the \citet{Fink2010a} models.
Even if some particular line-of-sight might compare more favourably to 
observed SNe~Ia than others (as does for example the southern line-of-sight
for the light curves in Figure~\ref{fig:m03_lightcurves}), all lines-of-sight 
should occur in nature -- including the most peculiar ones. Moreover,
the large dispersion in brightness at $B$-band maximum which we find in 
our model is in conflict with observations.

\subsubsection{Other models}
\label{sub:oo_los_all}

\begin{figure}
  \centering
  \includegraphics[width=8.5cm]{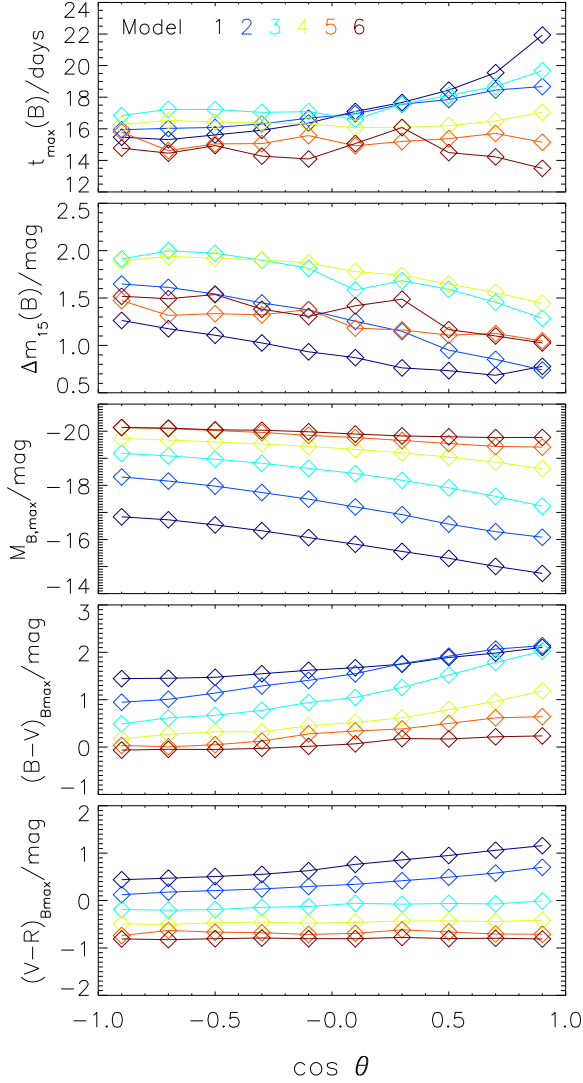}
  \caption{Light curve properties of the \citet{Fink2010a} models for 10 
    different viewing directions (different models are indicated by the 
    colour coding). From top to bottom the panels show the rise-time to 
    $B$-band maximum [$t_\mathrm{max}(B)$], the decline in $B$-band between 
    maximum light and 15 days thereafter [$\Delta m_{15}(B)$], the $B$-band 
    peak magnitude ($M_\mathrm{B,max}$), the $B-V$ colour at $B$-band maximum 
    [$(B-V)_\mathrm{Bmax}$] and the $V-R$ colour at $B$-band maximum 
    [$(V-R)_\mathrm{Bmax}$], respectively. The lines connecting the different
    data points are just to guide the eye.}
  \label{fig:los-dependence_all}
\end{figure}

Since the other models in the \citet{Fink2010a} sequence have the same
characteristic asymmetries, their light curves and spectra show a similar 
viewing-angle dependence. However, the strength of this viewing-angle 
dependence varies between the models, due to their different helium shell 
masses. To demonstrate this, Figure~\ref{fig:los-dependence_all} shows 
selected light curve properties of all models for the ten different 
viewing directions within our uniform grid in $\mu$.

As expected, the viewing-angle dependence decreases for smaller helium 
shell mass. Thus, we find a scatter of more than 2 magnitudes for the 
brightness at $B$-band maximum between the different lines-of-sight in 
Model 1 and 2, while Model 6 shows only a scatter of $\sim 0.3$ magnitudes. 
Similar trends can be observed for the $B-V$ and $V-R$ colours at $B$-band 
maximum. While Model 6 shows almost no dependence on the line-of-sight, 
the less massive models show a very clear trend for redder colours towards 
$\mu=0.9$ (as discussed for Model 3 in detail above).

For the rise times the situation is similar: while Models 1, 2 and 3 
(relatively massive shells) show a clear trend for increasing $B$-band 
rise times from $\mu=-0.9$ to $\mu=0.9$ due to the enhanced photon 
trapping by optically thick lines of iron-group elements, we do not 
observe a clear trend for Models 4, 5, and 6 (less massive shells). 
Recall, that the average rise times increase from Model 1 to 3, while 
Models 4 to 6 form a sequence of decreasing rise times. This was 
already discussed in Section~\ref{sub:oo_lcs}.

The enhanced trapping of photons by the iron-group layer of the northern 
hemisphere also affects the post-maximum decline rate in $B$ band. Thus,
we find decreasing values of $\Delta m_{15}(B)$ from $\mu=-0.9$ to 
$\mu=0.9$. This trend persists for all our models, although it is 
weaker for the models with the least massive shells.

\subsection{$\gamma$-ray emission}
\label{sub:oo_gamma}
\begin{figure}
  \centering
  \includegraphics[width=8.5cm]{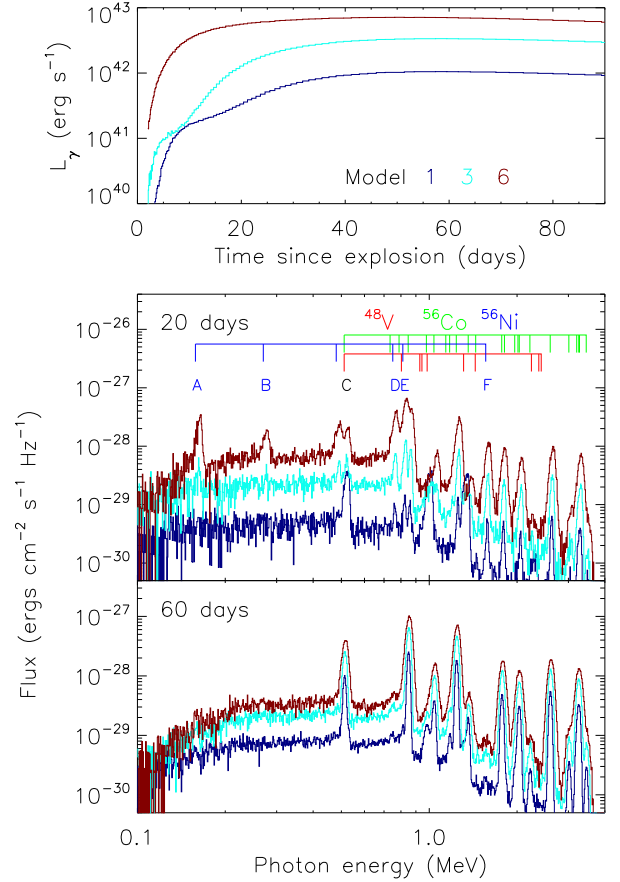}
  \caption{$\gamma$-ray light curves (top panel) and spectra (lower panels)
    for Models 1, 3 and 6 of \citet{Fink2010a} as indicated by
    the colour coding. The spectra shown are for two different epochs which
    correspond roughly to maximum light in $B$ band (20 days, middle panel)
    and in $\gamma$-rays (60 days, bottom panel). Line features discussed
    in the text are identified by labels (A -- F) in the middle panel (for their
    identification, see Table~\ref{tab:gammalineidentifiers}). In addition, the $^{56}$Ni, 
    $^{56}$Co and $^{48}$V line systems are indicated. The line systems of $^{48}$Cr, 
    $^{52}$Fe and $^{52}$Mn are omitted, since their life times are 
    sufficiently short that these nuclei have already decayed at the
    shown epochs. Note that at 20 days the line features in the spectra 
    are generally offset towards higher energies compared to the line 
    identifications, due to blue-shifted emission. At 60 days the emission
    comes from regions of lower velocity and the offset disappears.
    }
  \label{fig:gamma}
\end{figure}

Due to their peculiar composition including a mixture of the radioactive 
isotopes $^{56}$Ni, $^{52}$Fe and $^{48}$Cr close to the surface, 
$\gamma$-observations might provide an additional discriminant between
our sub-Chandrasekhar-mass models and more standard explosion models 
which do not show radioactive isotopes close to the surface. To investigate 
this, Figure~\ref{fig:gamma} shows $\gamma$-ray light curves and spectra 
for Models 1, 3 and 6 of our sequence. Line identifications are given for
some important features in Table~\ref{tab:gammalineidentifiers}.

\begin{deluxetable*}{ccl}
  \tabletypesize{\scriptsize}
  \tablecaption{Identification of $\gamma$-lines in Figure~\ref{fig:gamma}. \label{tab:gammalineidentifiers}}
  \tablehead{Identifier & Photon energy (MeV) & Source}

  \startdata
  A & 0.158 & $^{56}$Ni \\
  B & 0.270 & $^{56}$Ni \\
  C & 0.511 & Annihilation of positrons from $^{56}$Co and $^{48}$V\\
  D & 0.750 & $^{56}$Ni \\
  E & 0.812 & $^{56}$Ni \\
  F & 1.562 & $^{56}$Ni \\
  \enddata
  
\end{deluxetable*}

Broadly speaking, our $\gamma$-ray spectra are not dramatically 
different from those obtained by \citet{Sim2008a} for parameterised 
Chandrasekhar-mass models: they are dominated by strong emission lines, 
mainly due to $^{56}$Co, and a continuum which results from Compton 
scattering of line photons. 

Models 1 and 3 have characteristically similar $\gamma$-ray light curves.
After an initially fast rise (lasting for about 5 and 10 days after the 
explosion for Model 3 and 1, respectively) their light curves have a small 
plateau before passing to a second rise to maximum light at about 60 days.
In contrast, Model 6 shows no early rise/plateau but reaches maximum at
about 60 days in one continuous rise. These differences result from the 
significantly different masses of the helium shells of the models. Model~1 
has a rather massive shell of $0.126\,M_\odot$ with $\sim 0.02\,M_\odot$ 
of radioactive isotopes. The fast initial rise of the light curve of Model
1 is caused by $\gamma$-photons which originate from this outer shell.
Due to the small optical depth in the outer shell, those $\gamma$-photons 
start to stream freely at about 10 days and the emerging flux from the 
outer shell decreases. At the same time, however, $\gamma$-photons from 
the C/O core start to escape and keep the light curve rising until the 
core also becomes transparent to $\gamma$-rays at about 60 days after which
the $\gamma$-photons start to stream freely. 

For Model 3, we observe the same effect. However, due to its lower helium 
shell mass ($M_\mathrm{sh}=0.055\,M_\odot$) $\gamma$-photons from the shell 
escape even earlier (at about 5 days). Since the detonation of the C/O core 
of this model produces much more $^{56}$Ni, the $^{56}$Ni-rich region is 
far more extended than in Model 1. This means that $^{56}$Ni nuclei from the 
C/O core are present in regions where the Compton optical depth is relatively
low, allowing their $\gamma$-photons to more easily escape. Thus, photons 
from the core dominate the rise earlier than in Model 1. In the extreme 
limit of Model 6, the shell ($M_\mathrm{sh}=0.0035\,M_\odot$) is completely 
negligible and the massive $^{56}$Ni region in the C/O core extends so close 
to the surface that the rise of the light curve is totally dominated by 
$\gamma$-photons escaping from the C/O core.

A similar effect shows up in the early-time (20 days) $\gamma$-ray spectra 
of our models. While Model 6 shows a clear indication of the 0.158 and 
0.270 MeV lines of $^{56}$Ni, those are invisible for Model 1. Since the 
Compton cross-section increases with decreasing photon energy, soft-energy 
lines are most easily buried by photons being Compton down-scattered from 
higher energies and can only be observed if $^{56}$Ni is present at low 
optical depths \citep{Gomez1998a}. Thus the presence of the 0.158 and 
0.270 MeV lines of $^{56}$Ni in Model~6 is a direct consequence of the 
large extension of the $^{56}$Ni bubble in this model, compared to Model~1. 
The harder $^{56}$Ni lines at 0.750, 0.812 and 1.562 MeV, in contrast,
are also visible in Model~1. However, they are weaker due to the smaller
mass of $^{56}$Ni synthesized in Model 1. Note that despite containing 
$^{48}$Cr or $^{48}$V close to the surface, our models show no clear 
features of these radioactive isotopes in their $\gamma$-ray spectra.

An interesting effect concerns the 511 keV annihilation line: 
for Model 3 and 6 the strength of this line increases from 20 to 60 days 
significantly, but it does not for Model 1. In Model 1, the 511 keV line 
is dominated by positrons from $^{48}$V at 20 days. Being located in the 
outer layers, the annihilation photons escape easily making the line strong.
At 60 days most of the $^{48}$V has already decayed. Then, the 511 keV 
line results from the annihilation of $^{56}$Co positrons from the C/O 
core. For Model 3 and 6, in contrast, the line is always dominated by 
annihilation photons originating from $^{56}$Co positrons in the C/O 
core. Due to the longer life time of $^{56}$Co and the longer time it 
takes for photons to escape from the core, the strength of the 511 keV 
line increases from 20 to 60 days.

\section{Discussion} 
\label{sec:discussion}
In the last Section we presented synthetic observables for the 
sub-Chandrasekhar-mass double detonation models with minimum helium 
shell mass of \citet{Fink2010a}. Compared to observed SNe~Ia, these
models show some promising features:
\begin{enumerate}
  \item They predict a wide range of brightnesses that covers the 
    whole range of observed SNe~Ia. 
  \item Their light curve rise and decline times are for most bands
    in reasonable agreement with those observed. 
\end{enumerate}

But despite these positive features the \citet{Fink2010a} models
cannot account for all the properties of observed SNe~Ia since they 
have peculiar light curves and spectra:
\begin{enumerate}
  \item The colours of the models are too red compared to observed
    SNe~Ia. This is particularly obvious in the evolution of the 
    $B-V$ colour, where all the models are redder than 
    spectroscopically normal SNe~Ia for all epochs.
  \item The model spectra cannot reproduce the strong features of 
    intermediate-mass elements typical of SNe~Ia at maximum light.
\end{enumerate}

In detail, there are further problems concerning the exact light curve 
shapes and decline rates as well as a strong viewing-angle dependence 
which is caused by the point-like ignition of our models.

\begin{figure*}
  \centering
  \plotone{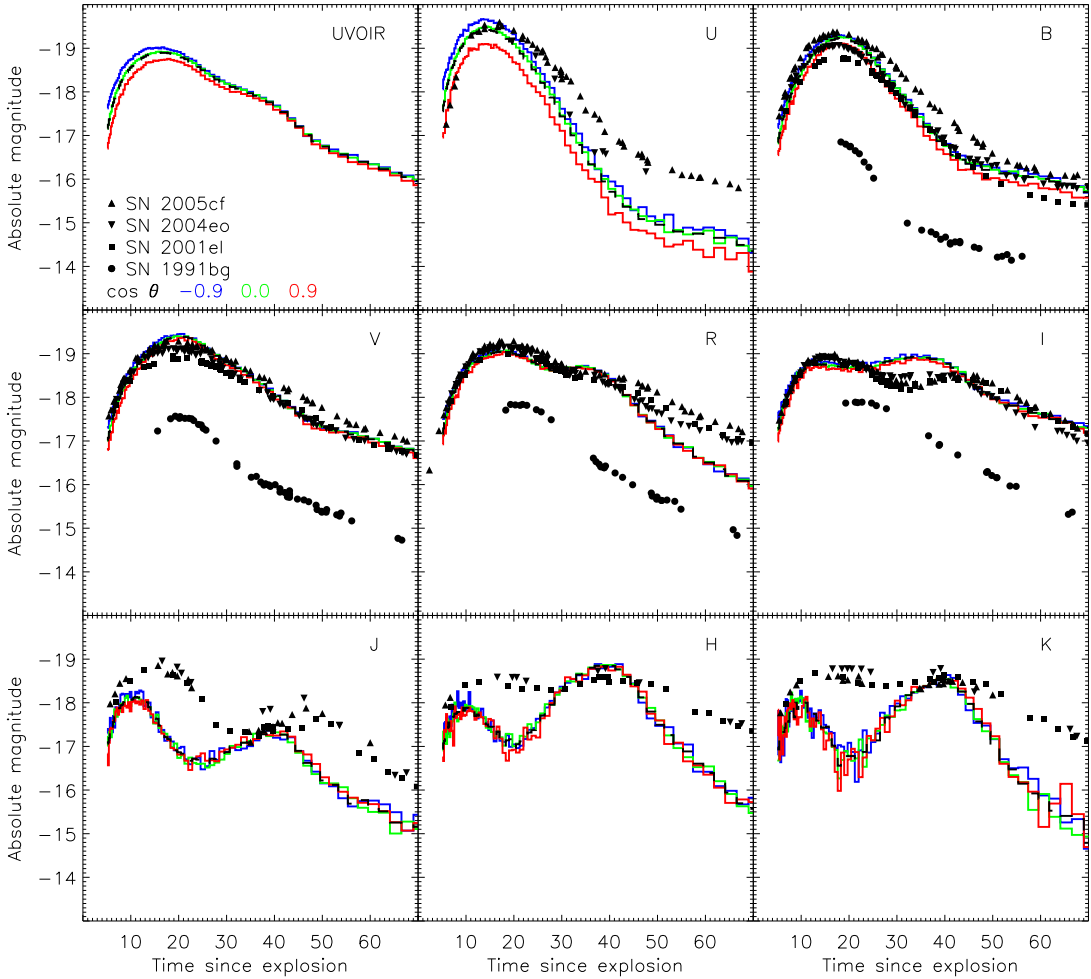}
  \caption{Selected line-of-sight dependent light curves of Model 3c as 
    indicated by the colour coding. For comparison angle-averaged
    light curves (black dashed) and photometrical data of our fiducial
    SNe 2005cf, 2004eo, 2001el and 1991bg (different symbols) are shown.}
  \label{fig:m03_nohe_lightcurves}
\end{figure*}

We have argued that all these problems are mainly due to the burning 
products of the helium shell. Moreover, \citet{Sim2010a}, have recently 
shown that detonations of centrally ignited spherically symmetric naked 
sub-Chandrasekhar-mass C/O WDs are capable of reproducing the observed 
diversity of light curves and spectra of SNe~Ia -- at least to a similar 
level of agreement as models of the more standard Chandrasekhar-mass 
delayed detonations. This naturally leads us to speculate on whether the 
helium shell in double detonation models could be altered in some way to 
reduce its negative impact on the synthetic observables.

\subsection{Influence of the helium shell}
\label{sub:disc_nohemodel}

We first explicitly investigate the extent to which the shortcomings 
and viewing-angle dependence of our models can indeed be attributed to 
the helium shell. For that purpose we constructed (for Model 3) a toy 
model which contains only the burning products of the detonation in 
the initial C/O core but not those of the initial helium shell. For 
the models of \citet{Fink2010a} this can be done in a straightforward
manner: since the models use two different sets of tracer particles to
simulate the nucleosynthetic yields of core and shell burning respectively
(see Section~3.3 of \citealt{Fink2010a}), we obtain such a model by
restricting our SPH-like reconstruction algorithm (Section~\ref{sec:rt})
to the core tracers. Properties of this ``core-only'' model (hereafter 3c)
are listed in Table~\ref{tab:modelparas}.

Figure~\ref{fig:m03_nohe_lightcurves} shows band-limited synthetic light 
curves obtained from our radiative transfer simulations for this model 
as seen equator-on and from the two polar directions. In contrast to the 
light curves of Model 3 in Figure~\ref{fig:m03_lightcurves}, which are 
strongly dependent on the viewing angle, the light curves of Model 3c show 
only a moderate line-of-sight dependence. Thus, at maximum light we find 
now only a variation of $\sim 0.5$ and $\sim 0.2$ magnitudes for $U$ and 
$B$ band, respectively. For Model 3 these values were significantly larger
($\sim3$ and $\sim2$ magnitudes, respectively). Redder bands show no 
significant line-of-sight dependence in Model 3c.

Moreover, the light curves of Model 3c give an excellent representation of 
SN~2004eo in the $B$, $V$ and $R$ bands. $U$ and $I$ are not in perfect 
agreement, but still reasonable compared to the agreement between other first 
principles explosion models and observed SNe~Ia (e.g. \citealt{Kasen2009a}). 
In particular, the colours of this toy model are now in good agreement 
with observed SNe~Ia and not too red as it is the case for Model 3. In 
the NIR bands, in contrast, the agreement is no better. However, the NIR 
light curves are much more difficult to model accurately since they require 
simulations with an extensive atomic data set to properly simulate flux 
redistribution by fluorescence which strongly affects these bands \citep{Kasen2006a,Kromer2009a}. 
Here, however, we have restricted ourselves for computational reasons to 
a simplified atomic data set (cd23\_gf-5 of \citealt{Kromer2009a}) with 
only $\sim 400,000$ lines. This has been shown to give reliable results in
the optical bands which, for our purposes, are the most important since
they are the most different between our toy Model 3c and Model 3 of
\citet{Fink2010a}. In the NIR bands, in contrast, Model 3c and Model 3 give
rather similar results (see Figure~\ref{fig:compare_m03_lightcurves}).

\begin{figure}
  \centering
  \plotone{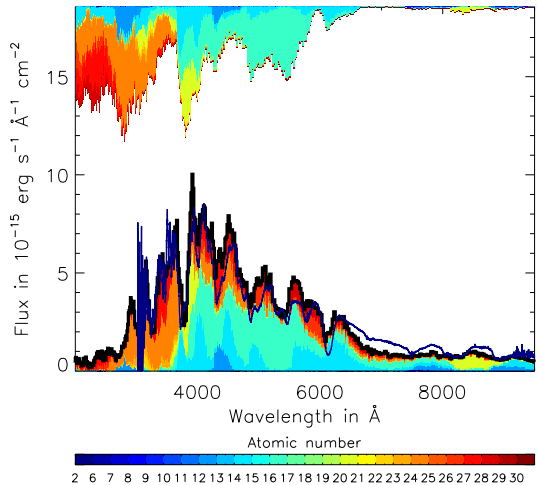}
  \caption{Angle-averaged (thick black line) spectra at three days before 
    $B$-band maximum for Model 3c. For comparison the blue line shows the 
    de-redshifted and de-reddened spectrum of SN~2004eo \citep{Pastorello2007b} 
    at the corresponding epoch. Note, that the flux is here in physical
    units and not scaled like in Figure~\ref{fig:maxspectra_all}. For a 
    description of the colour coding see Figure~\ref{fig:maxspectra_all}.}
  \label{fig:m03_nohe_spectrum}
\end{figure}

The good agreement between our toy Model 3c and observational data
does not only hold for band-limited light curves but also for individual 
spectral features as can be seen from Figure~\ref{fig:m03_nohe_spectrum} 
which shows a spectrum of Model 3c at 3 days before $B$-band maximum. 
Compared to SN~2004eo, our toy model succeeds in reproducing the 
characteristic spectral features of intermediate-mass elements in 
SNe~Ia. This is highlighted by our colour coding. Moreover, it shows 
an overall flux distribution which is in almost perfect agreement with 
the observed spectrum of SN~2004eo and we see no strong flux redistribution
by titanium (compare with Model 3 in Figure~\ref{fig:maxspectra_all}).

This confirms our conclusion from Section~\ref{sub:oo_los} that the 
peculiarities of our model spectra with respect to the observations
and their strong viewing-angle dependence are mainly due to the shell
material and its compositional asymmetries. It also shows that the 
off-centre ignition of the secondary detonation in the C/O core causes 
only a minor viewing-angle dependence which is on the order of the 
observed variation between SNe~Ia.

\subsection{Prospects}
\label{sub:disc_modifiedmdoel}
In light of the discussion above, we are motivated to speculate on how the 
influence of the helium shell might be reduced. In the sub-Chandrasekhar-mass
double detonation scenario, the helium shell cannot be removed entirely since 
it is required to trigger the detonation. Also the helium shell mass adopted
in the \citet{Fink2010a} models is already the minimum that might be expected
to detonate \citep{Bildsten2007a}. 
In Section~\ref{sec:oo}, however, we have argued that the differences
between our model spectra and observations are not a consequence of the
helium itself but of its particular burning products, namely titanium
and chromium produced in the outer layers. The yields of these elements 
are affected by details of the nucleosynthesis in the shell. 

The degree of burning in the shell material (and thus its final composition)
can be affected by the initial abundance of heavy nuclei (e.g.\ $^{12}$C) 
which in turn depend strongly on triple-$\alpha$ reactions during previous 
hydrostatic burning and dredge-up phases from the core \citep{Shen2009a}. 
Since the time-scale for $\alpha$-captures behind the detonation 
shock front is significantly shorter than that of triple-$\alpha$ reactions, 
such seed-nuclei can limit the $\alpha$-chain before reaching nuclear 
statistical equilibrium. If, for example, in a shell consisting of a mixture 
of $^{4}$He and $^{12}$C the number ratio of free $\alpha$-particles to 
$^{12}$C-nuclei on average is less than 6 (corresponding to a mass ratio 
of 2), the $\alpha$-chain will end at $^{36}$Ar. Thus, it is possible that 
more intermediate-mass elements and less titanium and chromium may be produced.
Therefore it is interesting to consider how the burning of the helium might 
be different from that found by \citet{Fink2010a} for different initial
compositions of the shell. A full study of this goes beyond the scope of 
this work and will be published in a follow-up study. Here, we illustrate 
the possibility of obtaining better agreement with data for just one 
example.

\begin{figure*}
  \centering
  \plotone{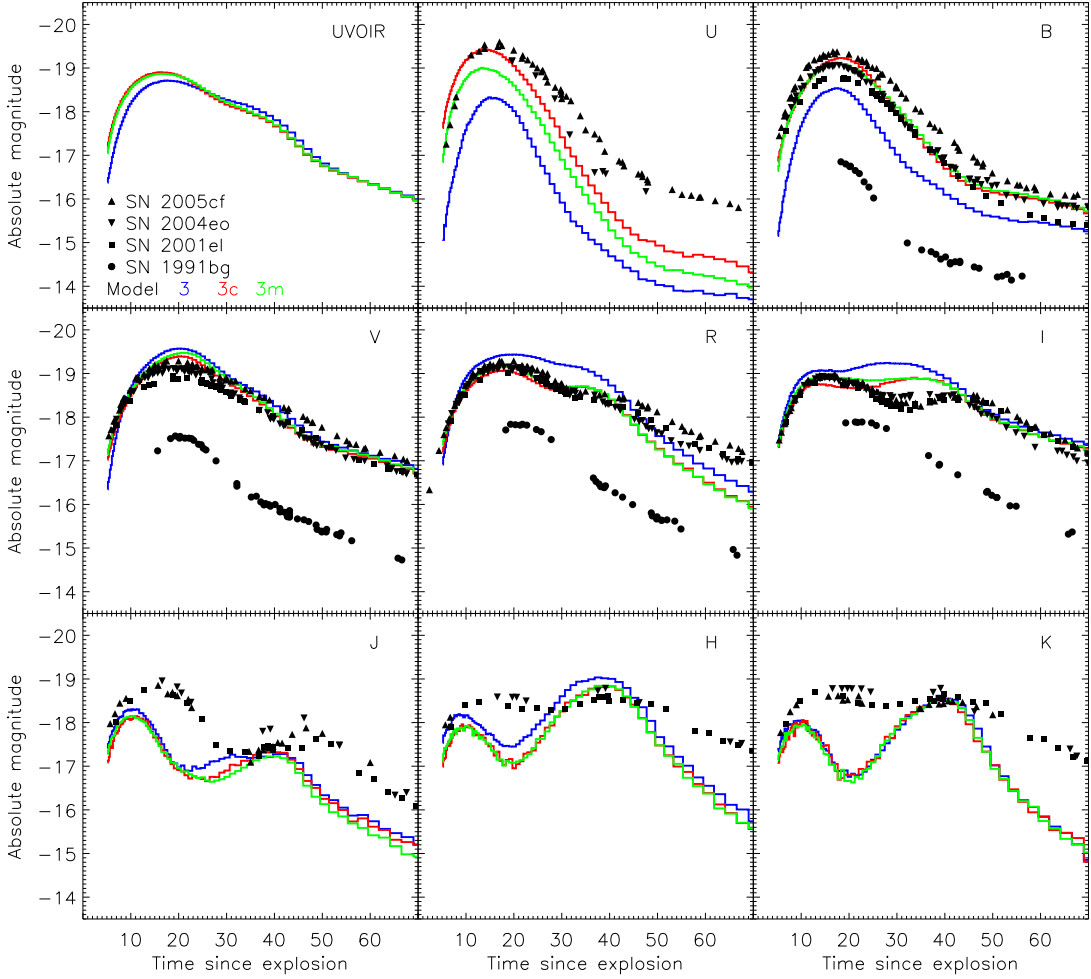}
  \caption{Angle-averaged $UVOIR$ bolometric and $U$,$B$,$V$,$R$,$I$,$J$,$H$,$K$ 
    band-limited light curves for Model 3, 3c and 3m of our model sequence 
    as indicated by the colour coding (compare Table~\ref{tab:modelparas} 
    for details on the models). For comparison angle-averaged
    light curves (black dashed) and photometrical data of our fiducial
    SNe 2005cf, 2004eo, 2001el and 1991bg (different symbols) are shown.}
  \label{fig:compare_m03_lightcurves}
\end{figure*}

In the \citet{Fink2010a} models it was initially assumed that the shell
consisted of pure helium. To demonstrate the sensitivity to the initial
composition of the shell, we set up another toy model. For this ``modified''
model (hereafter 3m), we homogeneously polluted the shell of Model 3 with 
34\% (by mass) of $^{12}$C and repeated the hydrodynamics and nucleosynthesis 
calculation (in the same way as described by \citealt{Fink2010a})\footnote{
We note, that the base temperature of the shell in Model 3m was decreased 
to $4\times10^8\,\mathrm{K}$ (compared to $6.7\times10^8\,\mathrm{K}$ in 
Model 3) to suppress further triple-$\alpha$ burning in the shell. As a
consequence the density and also the shell mass changes slightly compared 
to the original Model 3 (cf. Table~\ref{tab:modelparas}).}. 
We found that a core detonation was still triggered but the different 
shell burning led to a substantial reduction of the mass of $^{44}$Ti, 
$^{48}$Cr and $^{52}$Fe in the shell (nucleosynthetic yields for core 
and shell of the model are given in Table~\ref{tab:modelparas}).
Since the detonation tables of \citet{Fink2010a} are only valid for 
pure helium, Model 3m is not fully self-consistent. Nevertheless, it 
is a useful toy model to explore the basic effect of a modified shell 
composition.

Figure~\ref{fig:compare_m03_lightcurves} compares the angle-averaged
band-limited light curves of this modified model (3m), to those of Model 3
of \citet{Fink2010a} and our core-only toy model (3c). As can be seen, 
the modified Model 3m produces light curves very similar to those of 
Model 3c despite having about the same shell mass as Model 3. 
The most obvious difference between our modified and core-only models
occurs in the $U$ band: the titanium in the outer layer of Model 3m causes
some line blocking, leading generally to a dimmer $U$-band magnitude than 
for Model 3c which has no outer layer. Compared to Model 3, with its large 
titanium mass in the shell, however, this effect is much weaker. The 
$B$ band, which is strongly affected in Model 3, shows no significant 
titanium absorption for Model 3m. Another slight difference between Model 
3m and 3c occurs after the first peak in the $I$ band. Comparing the light 
curves of Model 3m to SN~2004eo, we find qualitatively similar agreement
as for Model 3c.

\begin{figure}
  \centering
  \plotone{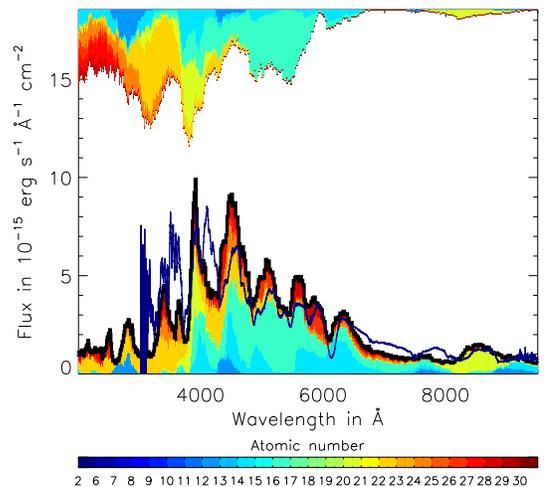}
  \caption{Angle-averaged (thick black line) spectra at three days before 
    $B$-band maximum for Model 3m. For comparison the blue line shows the 
    de-redshifted and de-reddened spectrum of SN~2004eo \citep{Pastorello2007b} 
    at the corresponding epoch. Note, that the flux is here in physical
    units and not scaled like in Figure~\ref{fig:maxspectra_all}. For a 
    description of the colour coding see Figure~\ref{fig:maxspectra_all}.}
  \label{fig:m03_modified_spectrum}
\end{figure}

This generally also holds for the angle-averaged spectrum at 3 days 
before $B$-band maximum, shown in Figure~\ref{fig:m03_modified_spectrum}.
Compared to Model 3c, where the agreement was almost perfect, there are 
some minor shortcomings. But the model is dramatically improved compared
to Model 3 of \citet{Fink2010a}. The small differences between Model 3c 
and 3m are again mostly due to the titanium in the outer layers which leads 
to pronounced absorption troughs bluewards of the Ca\,{\sc ii} H and K 
lines and redwards of $4,000\,\text{\AA}$. This suggests that Model 3m
still over-produced titanium in the shell. Interestingly, the enhanced 
calcium abundance in the outer layers (cf. Table~\ref{tab:modelparas}) 
leads to a stronger Ca\,{\sc ii} NIR triplet, bringing the model in better 
agreement with the spectrum of SN~2004eo at the corresponding epoch.
Therefore some \emph{calcium} in the outer shell is an \emph{improvement}
over Model 3c. This suggests that a slight further reduction in the
degree of burning so that titanium is further suppressed in favour of
calcium (just one step down the $\alpha$-chain) could lead to very good 
agreement.

In summary, polluting the initial helium shell of Model 3m with $^{12}$C 
significantly improved the agreement between our synthetic spectra and
light curves and those observed for SNe~Ia making this model a promising 
progenitor candidate for SNe~Ia. We stress again that this improvement 
results only from the change in the composition of the burning products 
of the helium shell which contains much less titanium and chromium for 
this model (the total shell mass stays about the same). Given that the 
initial composition of the helium shell depends on several processes 
including details of the accretion physics and hydrostatic burning phases 
that might precede the detonation or possible dredge-up of material from 
the C/O core, this leaves some scope to find sub-Chandrasekhar-mass 
double detonation models which give reasonable agreement with observed 
SNe~Ia. This, however, must be investigated by future follow-up studies 
that more fully explore the influence of the initial composition of the 
helium shell on the burning products and link the initial composition 
of the helium shell directly to the evolution of progenitor models. 
Moreover, we note that different ignition geometries might also lead 
to better agreement with observational data. In particular, more 
symmetric ignition geometries, e.g. ignition in an equatorial ring
or simultaneous ignition in multiple points (as studied by \citealt{Fink2007a}
for the case of more massive helium shells), are likely to alleviate 
the strong viewing-angle dependence found for the point-like ignition 
of the \citet{Fink2010a} models. 

Our results also highlight the strong sensitivity of the radiative transfer
to particular elements/ions (in our case titanium and chromium which 
represent only a tiny fraction of the ejecta mass yet dominate our
conclusions). This emphasizes the need for a better description of 
nuclear reaction rates and continued study of the radiative transfer 
processes (and atomic data) in order to quantify more fully the systematic 
uncertainties which arise due to the complexity of spectrum formation 
in supernovae. 
In particular, we note that almost all the flux redistribution done by 
titanium and chromium in our models is due to their singly ionized states. 
Since the current ionization treatment of {\sc artis} neglects non-thermal 
processes (see \citealt{Kromer2009a} for more details), we cannot 
exclude that the actual ionization state in the helium shell ejecta would 
be higher due to non-thermal ionization from the radioactive isotopes 
produced during the helium burning. This could also significantly improve 
the agreement between our models and observational data, as a numerical 
experiment with an artificially enforced higher ionization state for titanium 
and chromium has shown.

\section{Conclusion} 
\label{sec:conclusions}
In this paper we presented synthetic observables for the sub-Chandrasekhar-mass
double detonation models of \citet{Fink2010a}. 
We found that these models predict light curves which rise and fade on 
time-scales typical of SNe~Ia. Moreover, they produce a large range of 
brightnesses which covers the whole range of observed SNe~Ia. However, 
they do not account for all the properties of observed SNe~Ia since they 
have peculiar spectra and light curves. In particular, their $B-V$ colours 
are generally too red compared to observed SNe~Ia. This is in contrast to 
the results of earlier work on models with more massive helium shells 
\citep{Hoeflich1996b,Hoeflich1996c,Nugent1997a}. In addition, our model 
light curves and spectra show an unreasonably strong viewing-angle 
dependence due to the point-like ignition of the \citet{Fink2010a} 
models and the resulting ejecta asymmetries.

Detonation of a pure helium shell leads to a layer containing iron-group 
elements like titanium and chromium around the core ejecta. These
elements have a vast number of optically thick lines in the UV and 
blue part of the spectrum making them very effective in blocking the 
flux in these wavelength regions and redistributing it to the red.
We used a toy model to show that this layer of titanium and chromium
causes the peculiar red colours of our light curves and also the 
peculiar spectral features. Moreover, we found that this toy 
model reproduces the observed properties of SNe~Ia remarkably well.
The toy model also showed that the strong viewing-angle dependence 
of our models results from the compositional asymmetry in the helium
shell ejecta and not from the off-centre ignition of the C/O core. 
We stress, that the additional energy release in the shell, due to the 
production of radioactive nuclides during the helium burning, is relatively
inconsequential for our models -- even at $\gamma$-ray energies the
signatures of the surface $^{48}$Cr and $^{52}$Fe are not apparent. 
Instead, in the optical/UV the shell has a strong signature but this
is primarily due to the additional opacity in the outer layers which
affects the transport of energy from the core to the surface. We 
conclude that, if the double detonation sub-Chandrasekhar-mass model 
valid for normal SNe~Ia, the properties of the post-burning helium 
shell material need to be different from those in the \citet{Fink2010a} 
models.

Since \citet{Fink2010a} considered the limit of the least massive helium 
shells which might ignite a detonation in the helium shell, their models
represent the most optimistic case for reducing the influence of
the shell material by simply reducing the shell mass. However, we argue
that the mass of the helium shell ejecta is not the main problem but 
rather the peculiar composition including comparably large masses of 
titanium and chromium. We illustrated this using a second toy model,
where the initial composition of the helium shell was polluted with 34\% 
(by mass) of $^{12}$C. By providing additional seed-nuclei for 
$\alpha$-captures, this leads to burning products with lower atomic number 
(i.e.\ intermediate-mass elements rather than mainly iron-group elements). 
Spectra and light curves of this model which has about the same shell mass 
as Model 3 of \citet{Fink2010a} show comparably good agreement to observed 
SNe~Ia as the shell-less toy model.

Taking into account all these results, we argue that these systems might 
yet be promising candidates for SN~Ia progenitors. Much more work will 
be needed to properly investigate this possibility. Besides a more
detailed description of the excitation/ionization state in the radiative 
transfer modelling which includes non-thermal effects, we need a better 
understanding of the initial composition of the helium shell and the 
incomplete burning processes which take place in this material to reach
reliable predictions of the burning products of the helium shell. 
Although only a tiny fraction of the mass, the post-burning composition 
of the shell material is critical to assessing the viability of the 
sub-Chandrasekhar-mass double detonation scenario.

\section*{Acknowledgments}
We thank R.~Pakmor and S.~Taubenberger for many helpful comments and our
referee, David Branch, for a supportive report. The simulations presented 
here were carried out at the Computer Center of the Max Planck Society in 
Garching, Germany. This work was supported by the Deutsche 
Forschungsgemeinschaft via the Transregional Collaborative Research Center 
TRR 33 ``The Dark Universe'', the Excellence Cluster EXC153 ``Origin and 
Structure of the Universe'' and the Emmy Noether Program (RO 3676/1-1).

\bibliographystyle{apj}   

\end{document}